\documentclass[fleqn,usenatbib]{mnras}
\usepackage{fix-cm}

\usepackage{newtxtext,newtxmath}
\usepackage[T1]{fontenc}

\DeclareRobustCommand{\VAN}[3]{#2}
\let\VANthebibliography\thebibliography
\def\thebibliography{\DeclareRobustCommand{\VAN}[3]{##3}\VANthebibliography}

\usepackage{graphicx}	% Including figure files
\usepackage{amsmath}	% Advanced maths commands

\usepackage{hyperref}
\usepackage{booktabs}
\usepackage{amsmath}
\usepackage{ulem}
\usepackage{footmisc} % For \footref command
\usepackage{color}
\title[A Deep Look into the Intermediate-Age Open Cluster NGC 2506]{A Deep Look into the Intermediate-Age Open Cluster NGC 2506: What Binary Systems Reveal About Cluster Distance and Age}

\author[K. Yakut et al.]{
Kadri Yakut,$^{1,2}$\thanks{E-mail: kadri.yakut@ege.edu.tr(KY)}
Belinda Kalomeni$^{1}$
Saul Rappaport$^{3}$
and
Veselin Kostov$^{4,5}$
\\
$^{1}$Department of Astronomy and Space Sciences, Faculty of Science, Ege University, 35100, {\.I}zmir, Türkiye\\
$^{2}$Institute of Astronomy, The Observatories, Madingley Road, Cambridge CB3 OHA, UK\\
$^{3}$Department of Physics, Kavli Institute for Astrophysics and Space Research, M.I.T., Cambridge, USA\\
$^{4}$NASA Goddard Space Flight Center, Greenbelt, MD 20771, USA\\
$^{5}$SETI Institute, 189 Bernardo Ave, Suite 200, Mountain View, CA 94043, USA}
\date{Accepted XXX. Received YYY; in original form ZZZ}

\pubyear{\the\year{}}

\begin{document}
\label{firstpage}
\pagerange{\pageref{firstpage}--\pageref{lastpage}}
\maketitle

\begin{abstract}
Using high-precision observations from the space-based \textit{Gaia} and \textit{TESS} missions, complemented by ground-based spectroscopic data and multi-band photometric surveys, we perform a detailed investigation of the Galactic open cluster NGC~2506. We present a new analysis of the intermediate-age open cluster NGC~2506, using joint fits to the radial velocities (RVs) and spectral energy distributions (SEDs) of five double-lined binary systems, including two eclipsing binaries. The analysis yields self-consistent estimates of the cluster's age, distance, and extinction, based on 18 free parameters: 10 stellar masses, 5 orbital inclinations, and common values for age, distance, and $A_V$. The SED fitting incorporates stellar isochrones, and the resulting parameters are examined through HR diagrams (R--$T_{\rm eff}$, R--M, and M--$T_{\rm eff}$) to assess evolutionary consistency. The age we derive for the cluster is $1.94 \pm 0.03$ Gyr for an assumed [Fe/H] = -0.30, and a fitting formula is given for extrapolation to other metallicities.  The distance we find from the SED fitting is $3189 \pm 53$ pc, and this is to be compared with our own inference from the Gaia data which is $3105 \pm 75$ pc, based on 919 stars identified as cluster members. Our results demonstrate the power of binary systems in tightly constraining cluster-wide age and distance at this evolutionary stage. This approach represents one of the most accurate characterizations of an intermediate-age open cluster using multiple binary systems.
\end{abstract}

\begin{keywords}
Binary stars --- Fundamental parameters of stars ---  (Galaxy:) open clusters and associations: individual (NGC 2506) --- (Galaxy:) open clusters and associations: general ---  proper motions
\end{keywords}

%%%%%%%%%%%%%%%%%%%%%%%%%%%%%%%%%%%%%%%%%%%%%%%%%%

%%%%%%%%%%%%%%%%% BODY OF PAPER %%%%%%%%%%%%%%%%%%

\section{Introduction} % Section starts here
\label{sec:intro}

The open cluster NGC~2506, located in the anti-centre direction of the Galaxy, is a well-populated, metal-poor, intermediate-age cluster that has long attracted attention due to its importance for understanding both stellar evolution and Galactic chemical evolution. Its colour--magnitude diagram (CMD) exhibits a well-defined main sequence and an extended giant branch morphology, coupled with low reddening and clear evidence of mass segregation. These features make NGC~2506 a valuable laboratory for testing stellar evolution models. As noted by \citet{chen2004}, the cluster's elongated morphology and central concentration suggest it is dynamically evolved. With an age of approximately 2\,Gyr and a distance of $\sim$3\,kpc, NGC~2506 occupies a transitional region of the Galactic disc, both chemically and structurally. The presence of diverse stellar populations, including main-sequence stars, giants, blue stragglers, yellow giants, and white dwarfs, further enhances its astrophysical significance.

The chemical composition of NGC~2506 has been extensively studied using both photometric and spectroscopic techniques across a range of resolutions. Early estimates of its iron abundance based on photometry include [Fe/H]~$= -0.43$ by \citet{McClure1981}, [Fe/H]~$= -0.52$ by \citet{friel1993}, [Fe/H]~$= -0.48$ by \citet{piatti1995}, and [Fe/H]~$= -0.58$ by \citet{geisler1992}. A recalibration of David Dunlap Observatory (DDO) photometry by \citet{Anthony-Twarog2016} yielded [Fe/H]~$= -0.37$. High-resolution spectroscopic studies have provided more precise abundance measurements: \citet{Carretta2004} reported [Fe/H]~$= -0.24 \pm 0.09$ based on four red giants, and this value reduces to $-0.20 \pm 0.01$ when only two stars are considered. \citet{mikolaitis2011} found [Fe/H]~$= -0.24 \pm 0.05$ from the same stars and extended their analysis to 26 elements, observing reduced C/N and $^{12}$C/$^{13}$C ratios in clump stars, which suggest extra mixing effects potentially triggered by helium flash episodes. \citet{reddy2012} derived [Fe/H]~$= -0.19 \pm 0.06$ from three giants. These metallicity determinations, as are the following ones, summarized in Table \ref{tab:ngc2506_full}.

Photometric determinations support these spectroscopic findings. Using $uvbyCaH\beta$ CCD photometry, \citet{Anthony-Twarog2016, Anthony-Twarog2018} derived [Fe/H]~$= -0.296 \pm 0.011$ from $m_1$ and H$\beta$, and [Fe/H]~$= -0.317 \pm 0.004$ from $hk$ and H$\beta$, yielding a weighted average of $-0.316 \pm 0.033$. These values were also supported by ANNA spectroscopic parameters. \citet{Knudstrup2020} found [Fe/H]~$= -0.36 \pm 0.10$ and [$\alpha$/Fe]~$= 0.10 \pm 0.10$ from an analysis of detached binaries and red giant branch (RGB) stars. In this study, Mg, Si, and Ca show mild $\alpha$-enhancement, while Ti does not. These values were also adopted by \citet{Nardiello2023}. Regarding light elements, \citet{Anthony-Twarog2018} highlighted lithium depletion on the subgiant and giant branches and reported the discovery of a Li-rich giant star (NGC~2506~4128), possibly linked to post-helium flash production. 
A correlation between rotational velocity and lithium abundance was also noted, indicating depletion below a threshold projected rotational velocity (V$_{\rm rot} \equiv v \sin i$), typically estimated from spectroscopic line broadening \citep{Anthony-Twarog2018}.

Determining the age of NGC~2506 is crucial for constraining the evolutionary states of its stars and for interpreting Galactic disc formation and evolution. Age estimates have varied over time, ranging from photometric isochrone fitting to detailed spectroscopic and binary modelling. \citet{salaris2004} proposed an age of $2.14 \pm 0.35$\,Gyr, while \citet{xin2005} suggested a higher value of 3.4\,Gyr. 
Based on VI CCD photometry and theoretical Padova isochrones, \citet{Lee2012} derived an age of $2.31 \pm 0.16$,Gyr for NGC 2506 through $\chi^2$ minimisation of the colour–magnitude diagram.
In contrast, \citet{Anthony-Twarog2016} reported $1.85 \pm 0.05$\,Gyr using Victoria-Regina models and $uvbyCaH\beta$ photometry. These studies demonstrate the age sensitivity of features such as the giant clump and turn-off morphology. Given its moderate metal deficiency ([Fe/H]~$\sim -0.3$), NGC~2506 represents a well-defined evolutionary test case.

Detached eclipsing binaries (DEBs) offer one of the most accurate age-dating methods via mass–radius comparisons. \citet{Knudstrup2020} modelled three DEBs (WOCS 5002, WOCS 17003 and NGC 2506 ADG V5) along with RGB stars and determined an age of $2.01 \pm 0.10$\,Gyr using BaSTI isochrones. Their results emphasise the necessity of incorporating convective core overshooting into the models. \citet{Panthi2022} analysed the spectral energy distributions (SEDs) of blue and yellow straggler stars, confirming an age consistent with 2.2\,Gyr. \citet{Nardiello2023}, using JWST data, adopted the same age and chemical composition from \citet{Knudstrup2020}. These values also align with other cluster properties, including dynamics \citep{Lee2012, Lee2013}, lithium depletion \citep{Anthony-Twarog2018}, and variable star populations such as $\delta$~Scuti and $\gamma$~Doradus stars \citep{Arentoft2007}.

The distance to NGC~2506 has been determined with increasing precision through both photometric and astrometric methods. \citet{Lee2012} derived a true distance modulus of $(V - M_V)_0 = 12.47 \pm 0.08$ with $E(B-V) = 0.03 \pm 0.04$. \citet{Anthony-Twarog2016} found $(m - M) = 12.75 \pm 0.10$ and $E(B-V) = 0.058 \pm 0.001$ using $uvbyCaH\beta$ photometry. \citet{Marconi1997} estimated the distance modulus to lie between 12.5 and 12.7. These values correspond to distances in the range 3100–3300\,pc, consistent with modern results.

Gaia-based astrometry has significantly refined distance estimates. \citet{Knudstrup2020} calculated a distance of $3101 \pm 170$\,kpc from Gaia parallaxes, while \citet{Gao2020} derived $3111 \pm 21$\,pc with a distance modulus of $12.464 \pm 0.014$. These values have been adopted by \citet{Panthi2022} and \citet{Nardiello2023}, the latter confirming $3100 \pm 175$\,pc from JWST--Gaia cross-matching and reporting a modulus of $12.46 \pm 0.12$. These modern determinations are substantially lower and more precise than earlier estimates such as 3460\,pc \citep{Arentoft2007} and 3880\,pc \citep{Rangwal2019}.

Binary systems within NGC~2506 have played a crucial role in constraining the cluster's age and distance. By analyzing three detached eclipsing binaries (DEBs) and four red giant branch stars in NGC 2506,  \citet{Knudstrup2020} determined precise stellar parameters. The system WOCS 5002, being in a critical evolutionary phase for age-dating, provided the tightest constraints, with component properties of: $M_1 \approx 1.52 M_\odot$, $M_2 \approx 1.50 M_\odot$, $R_1 \approx 3.1 R_\odot$, $R_2 \approx 2.4 R_\odot$, $T_{\text{eff,1}} \approx 6560$ K, and $T_{\text{eff,2}} \approx 7100$ K. These values were derived in conjunction with detailed orbital solutions, including eccentricities and third-body indicators. Subsequent work by \citet{Linck2024} expanded this to six SB1/SB2 systems using Gaia EDR3 astrometry and multi-epoch radial velocities, revealing a number of yellow straggler candidates and highlighting the highly eccentric ($e \sim 0.6$) nature of system WOCS 5002. Combined with light curve and SED analyses, these binary systems provide a robust framework for probing the evolutionary state of the cluster.

A summary of representative values for the metallicity, age, reddening, and distance of NGC~2506 as reported in the literature is provided in Table~\ref{tab:ngc2506_full}. While there is general agreement that the cluster is of intermediate age ($\sim$2~Gyr) and moderately metal-poor ($\rm [Fe/H] \approx -0.3$), the reported values vary non-negligibly across different studies. The mean metallicity from the literature is $-0.30 \pm 0.05$~dex, with age estimates ranging from 1.5 to 3.4~Gyr. Similarly, values for $E(B-V)$ span 0.04 to 0.10~mag, and the derived distances vary by more than 500~pc. These ranges highlight the persistent uncertainties in the fundamental parameters of the cluster and underscore the need for a comprehensive, self-consistent approach.

Building on the availability of high-quality astrometric, spectroscopic, and photometric data, we analyze the intermediate-age open cluster NGC 2506 by leveraging five double-lined spectroscopic binaries, two of which are eclipsing. Through a combination of radial velocity measurements, the \textit{Transiting Exoplanet Survey Satellite} (TESS) \citep{Ricker2015} light curves, \textit{Gaia} DR3 \citep{Gaia_Collaboration2023} astrometry and SED modeling we provide refined estimates of the cluster’s fundamental parameters without relying on traditional main-sequence fitting techniques. Stellar evolution models are used to simultaneously determine individual component properties and the global cluster age and distance. Additionally, we incorporate astrometric information from \textit{Gaia} DR3 to independently validate the derived cluster distance and membership. This multi-faceted approach enables a high-precision characterization of NGC 2506, shedding light on the stellar evolution status of its binary population. The approach we use here is similar to what we did in the study of NGC 188 \citep{Yakut2025}.

\begin{table*}
\centering
\caption{Literature estimates of [Fe/H], age, reddening, and distance for NGC~2506 based on a variety of methods. The mean and rms scatter of literature values are given, along with the parameters derived in this study for comparison.}
\label{tab:ngc2506_full}
\begin{tabular}{llllll}
\hline
[Fe/H]           & Age            & E(B--V)          & Distance / DM    & Method                    & Reference \\
dex              & Gyr            & mag              & pc/ mag          &                           &  \\
\hline
                 &                & $0.10$           & 12.0                & CMD+Photometry            & \citet{Purgathofer1964} \\
$-0.41 \pm 0.11$ &                & $0.10$           & $11.8$           &                           & \citet{Janes1979} \\
                 & $3.4$          & 0.05                 & $12.2$           &                           & \citet{McClure1981} \\
$-0.5$           & $1.5 - 2.2$    & $0-0.07$         & $12.6 \pm 0.1$   & CMD+Photometry            & \citet{Marconi1997} \\
$-0.20 \pm 0.01$ & $2.0$          & $0.073 \pm 0.009$&                  & High-res spectroscopy     & \citet{Carretta2004} \\
                 & $1.79$         &                  & $3421$           & CMD+Photometry            & \cite{Henderson2007} \\
$-0.24 \pm 0.06$ & $2.31 \pm 0.16$& $0.057 \pm 0.008$& $3195 \pm 170$   & CMD + Iso (VI Phot.)      & \citet{Lee2012} \\
$-0.27 \pm 0.05$ & $1.85 \pm 0.05$& $0.058 \pm 0.001$& $12.75\pm 0.10$  & uvbyCaH$\beta$ Photometry & \citet{Anthony-Twarog2016} \\
$-0.29 \pm 0.03$ & $2.09 \pm 0.14$& $0.06 \pm 0.02$  & $3880 \pm 420$   & Gaia DR2 CMD + Iso        & \citet{Rangwal2019} \\
$-0.36\pm 0.10$  & $2.0 \pm 0.1$  & $0.05$           & $3100 \pm 170$    & EB + CMD                  & \citet{Knudstrup2020} \\
$-0.27$          &                &                  & $3111 \pm 21$    & CMD                       & \citet{Gao2020} \\
$-0.32$          & $2.3$          & $0.06$           & $3200$           & AGB + Gaia DR3            & \citet{Marigo2022} \\
                 &                & $0.05$           &                  & Gaia DR3 Photometry       & \citet{Nardiello2023} \\
                 & $2.2 \pm 0.2$  &                  &                  & Blue Straggler + UVIT     & \citet{Vaidya2024} \\
$-0.28$          &                & $0.04$           & $3000$           & CMD + synthetic fitting   & \citet{Linck2024} \\
\hline
$-0.30 \pm 0.05$ & $2.14 \pm 0.35$ & $0.062 \pm 0.012$ & $3322 \pm 325$ & Literature Mean (rms)     & -- \\
\hline
$-0.30$          & $1.94 \pm 0.03\dag$ & $0.068 \pm 0.005$& $3189 \pm 53$    & Binary + SED             & This study\\
\hline
\end{tabular}
\smallskip

\textbf{Notes:} Methods: EB = Eclipsing Binary, CMD = Color–Magnitude Diagram, Iso = Isochrone fitting, SED = Spectral Energy Distribution, Spec = Spectroscopy, AGB = Asymptotic Giant Branch, Gaia = Gaia XP photometry,  DM = Distance Modulus. $^\dag$ Statistical uncertainty. 
\end{table*}

%%%%%%%%%%%%
%%%SECTION
%%%%%%%%%%%%
%\section{Observations, data analysis and modelling}
\section{Observations}
\label{sec:obs}

\begin{table*}
\scriptsize
\caption{Astrometric and photometric properties of the five double-lined binary systems analyzed in this study, including Gaia DR3 and TESS IDs, proper motions, parallaxes, and Gaia colours.}
\label{tab:basic_parameters_all}
\begin{tabular}{l r r r r r }
\hline
                                & WOCS 5002           &  WOCS 17003       & WOCS 24005          &  WOCS 17014         & WOCS 9004           \\
                                & A                   & B                 &   C                 &    D                &     E            \\
\hline
TIC ID                          & 169660646           & 169716051          & 169715921          & 169660231           & 169660719          \\
Gaia DR3 ID                     & 3038044880608377216 & 3038045567803107072& 3038046495516032640& 3038042990822882560 &3038046358077120256 \\
Alias                           & V2032, YSS1         & V943 Mon, V4       &                    &                     &                 \\
$\alpha$ ($^{\rm o}$)           & 120.002             & 120.0343           & 120.024            & 119.970             & 119.999          \\
$\delta$ ($^{\circ}$)           & -10.761             & -10.764            & -10.733            & -10.874             & -10.746            \\
$\mu_{\alpha}$ (mas~yr$^{-1}$)  & -2.650              & -2.612             & -2.510             & -2.590              & -2.524           \\
$\mu_{\delta}$ (mas~yr$^{-1}$)  &  3.939              & 3.906              & 3.994              & 3.917               & 3.899            \\
$\varpi$ (mas)                  & 0.2438              & 0.2812             & 0.2939             & 0.2515              & 0.2496             \\
$G$ (m)                         & 13.628              & 14.561             & 15.233             & 15.265              & 14.250           \\
$G_{\rm{BP}}-G_{\rm{RP}}$ (mag) & 0.611               & 0.595              & 0.614              & 0.613               & 0.560              \\
\hline
\end{tabular}
\end{table*}

The open cluster NGC~2506, located in the Galactic anti-centre direction and known for its intermediate age and moderately low metallicity, has been the focus of numerous photometric and spectroscopic studies. For the present work, we identified five double-lined spectroscopic binary (SB2) systems within the cluster that are confirmed members based on radial velocity and astrometric criteria derived from Gaia~DR3. Among these, two systems (WOCS~5002 and WOCS~17003) show clear photometric eclipses, while the remaining three (WOCS~9004, WOCS~17012, and WOCS~24005) are detached, non-eclipsing SB2 binaries. 
The selection was made to avoid systems affected by prior mass transfer, to exclude known blue stragglers, and to ensure the targets had sufficient brightness to allow for accurate SED analysis. Naturally, the most important criteria are that the selected binary systems exhibit clearly double-lined radial velocities, possess high signal-to-noise Gaia DR3 astrometry, and have well-sampled RV curves with sufficient wavelength coverage for broadband SED analysis. Instead of including all SB2 systems listed in the literature (e.g., Table~7 of \cite{Linck2024}), we focused on selecting only those binaries that were best suited for our joint RV+SED analysis framework. The systems span a range of orbital periods from approximately 2.4 to 73 days. These five binaries were jointly modeled to derive their orbital and stellar parameters and to provide independent constraints on the age and distance of NGC 2506.

Radial velocity data for the selected binary systems were obtained from two primary sources (see Figure~\ref{fig:RVs} for the observed RV variations as a function of orbital phase). The first is the extensive spectroscopic survey by \citet{Knudstrup2020}, who derived orbital parameters for several eclipsing binaries and red giant stars in NGC~2506. The second is the more recent work by \citet{Linck2024}, which provides refined orbital solutions for multiple SB1 and SB2 systems using multi-epoch spectroscopy obtained with ground-based telescopes such as WIYN and the MMT Observatory, combined with Gaia EDR3 astrometry. For all five targets, we re-analysed the RV data using a custom Markov Chain Monte Carlo (MCMC)-based pipeline to derive consistent orbital solutions and uncertainty estimates.

Complementary photometric data were obtained from TESS, which observed NGC~2506 in multiple sectors. We extracted 2-minute cadence light curves for the eclipsing binaries using the \texttt{Lightkurve} package \citep{Cardoso2018}, and also accessed publicly available products from the TESS-Gaia Light Curve (TGLC) archive \citep{TGLCref2023} for comparison and validation.

Astrometric information from Gaia~DR3 was used both for membership selection (based on proper motion and parallax) and as input to the SED construction through the inclusion of Gaia G, B$_P$, and R$_P$ magnitudes. In addition, multi-band photometry from surveys including 2MASS \citep{2MASSS2006}, Pan-STARRS \citep{PanSTARS2002}, WISE \citep{WISE2010}, SDSS \citep{SDSS2000}, and Gaia \citep{Gaia_Collaboration2023} was compiled via the VizieR SED tool \citep{ochsenbein00} to construct broadband SEDs covering a wavelength range from 0.35 to 4.6~$\mu$m. These photometric datasets were jointly analysed with the spectroscopic and light curve data to determine the fundamental stellar parameters of the binary components and to estimate the cluster's age, distance, and extinction. A summary of the key observational properties of the selected systems is provided in Table~\ref{tab:basic_parameters_all}.

\section{Light and radial velocity curve modeling }
\label{sec:lcrvmodel}

\begin{figure*}
\centering
\includegraphics[width=0.45\linewidth]{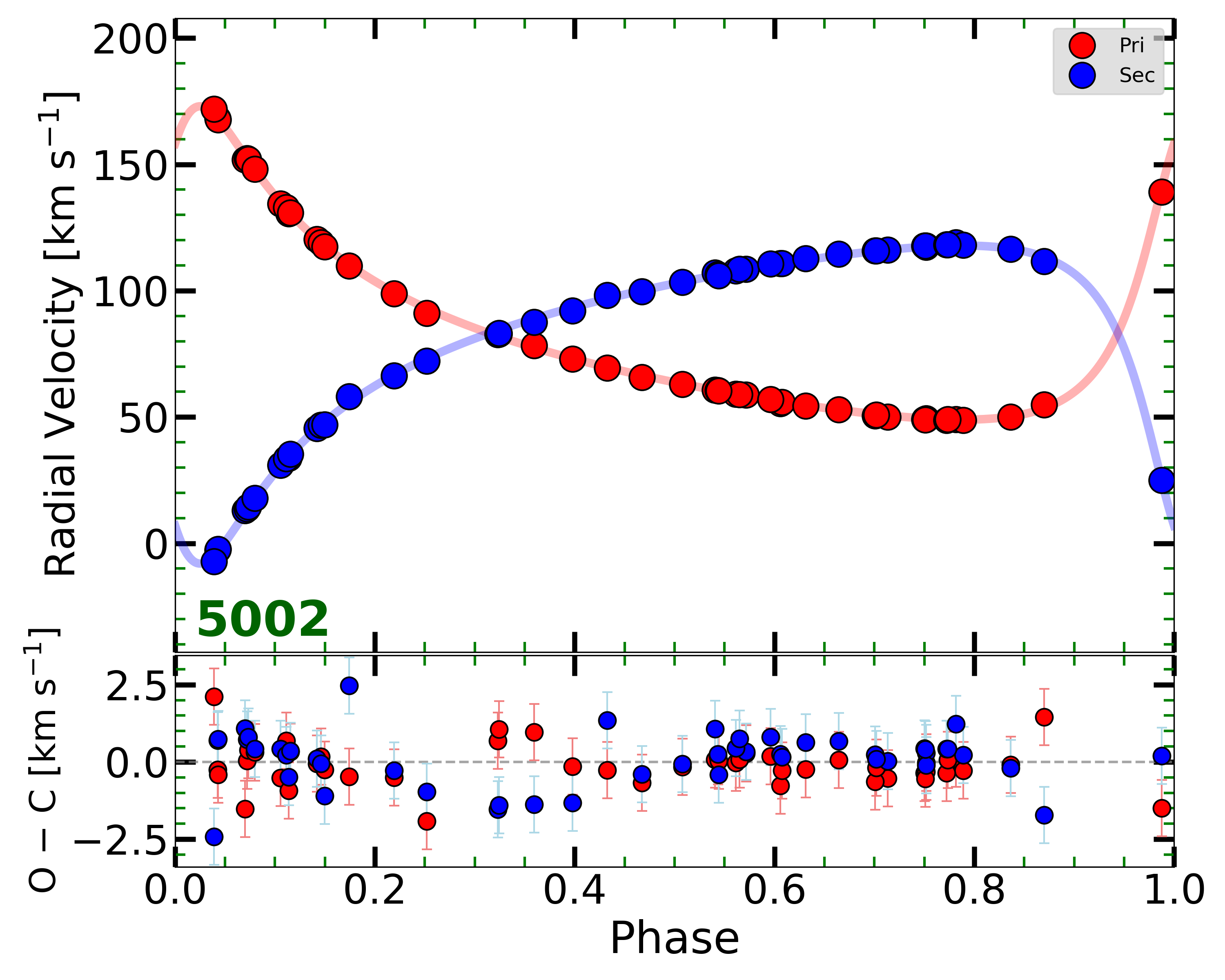}
\includegraphics[width=0.45\linewidth]{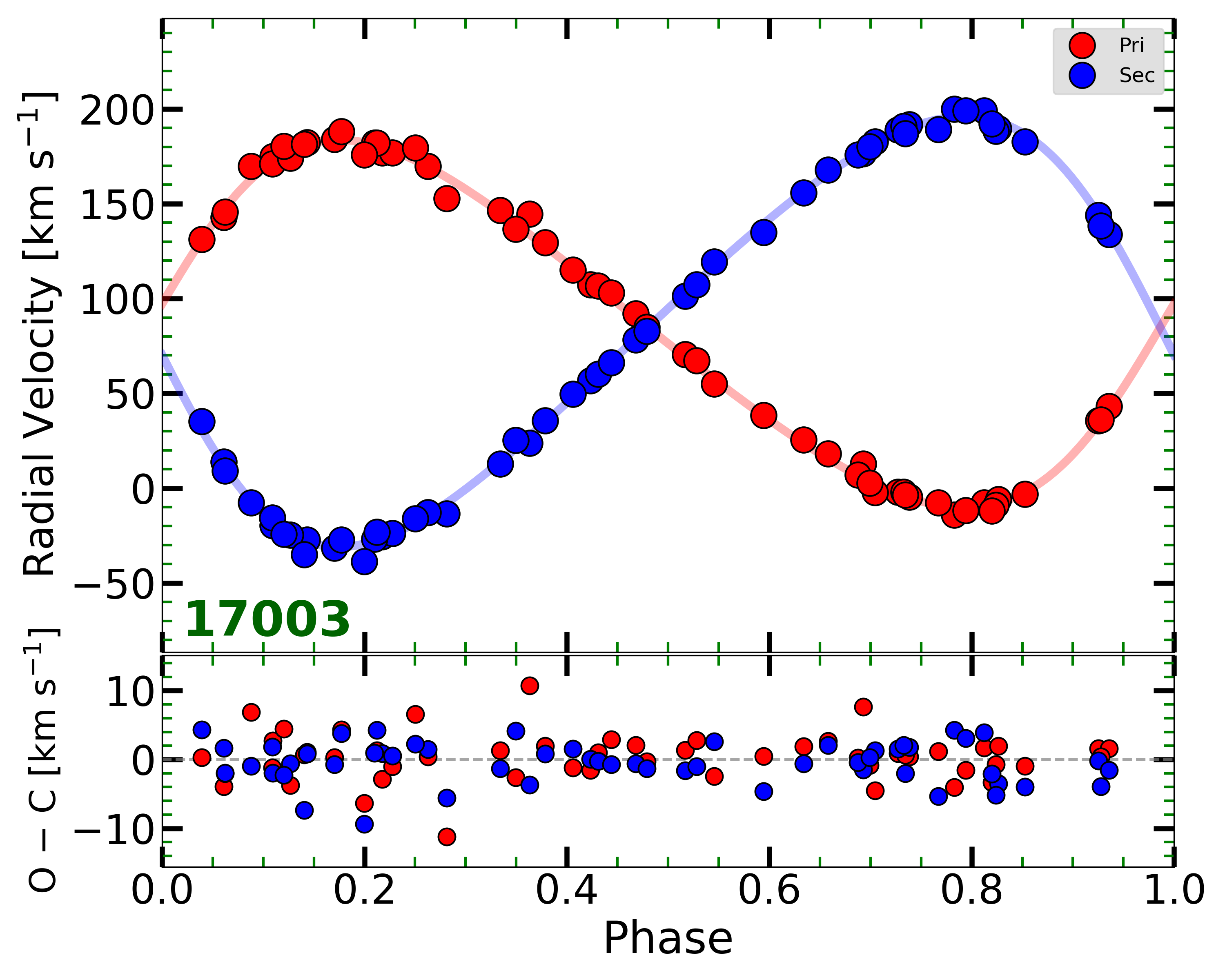}\\
\vspace{2mm}
\includegraphics[width=0.45\linewidth]{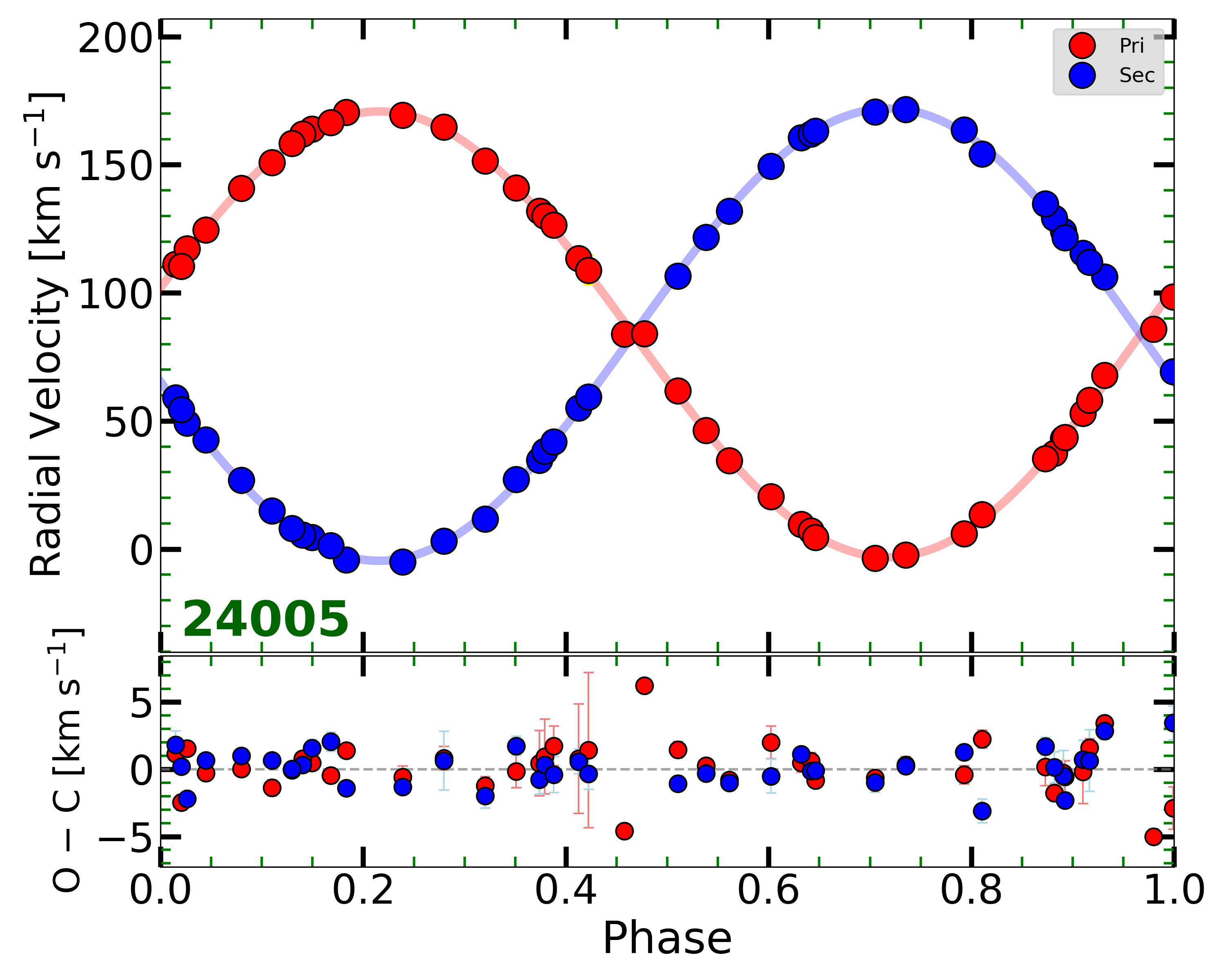}
\includegraphics[width=0.45\linewidth]{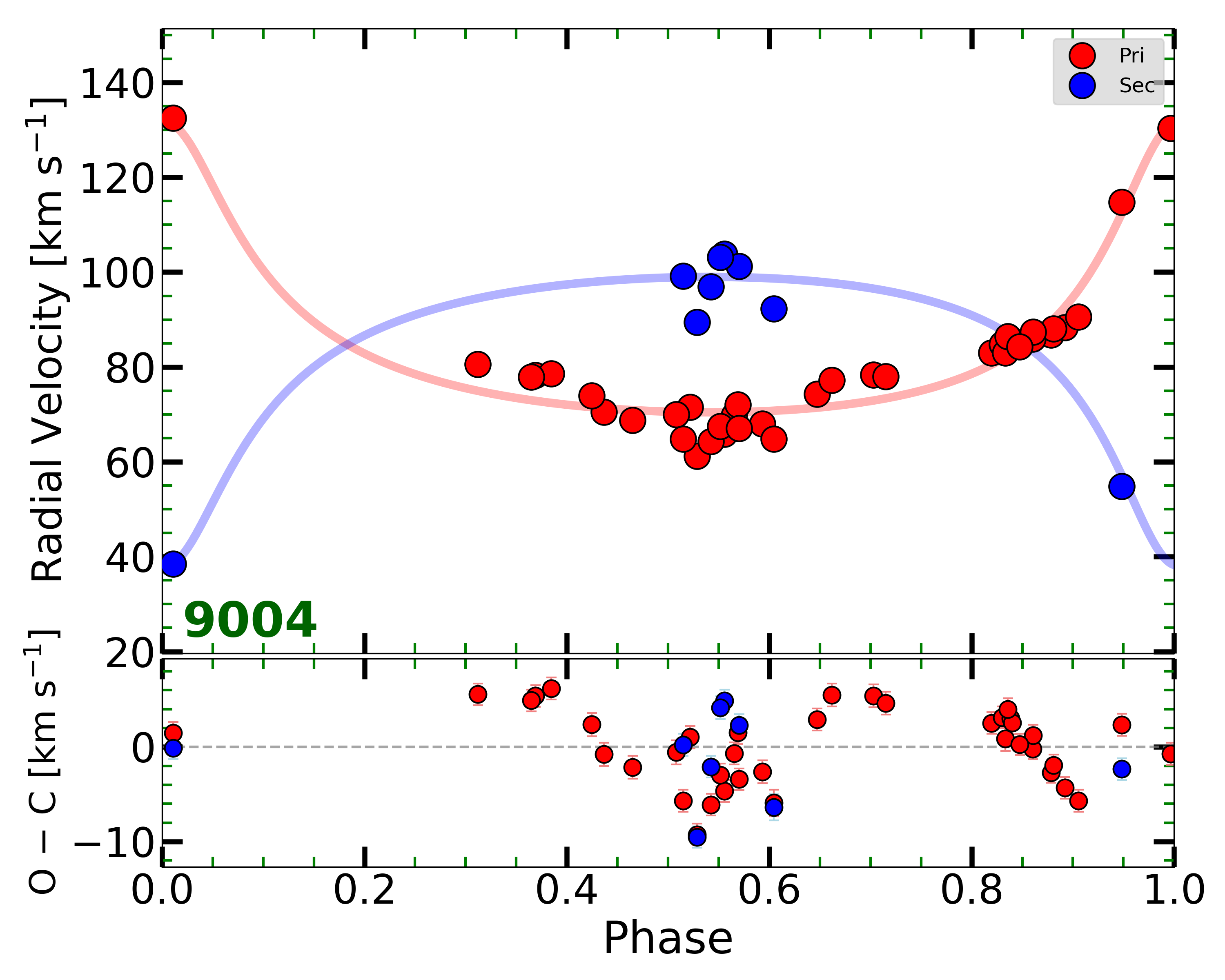}\\
\vspace{2mm}
\includegraphics[width=0.45\linewidth]{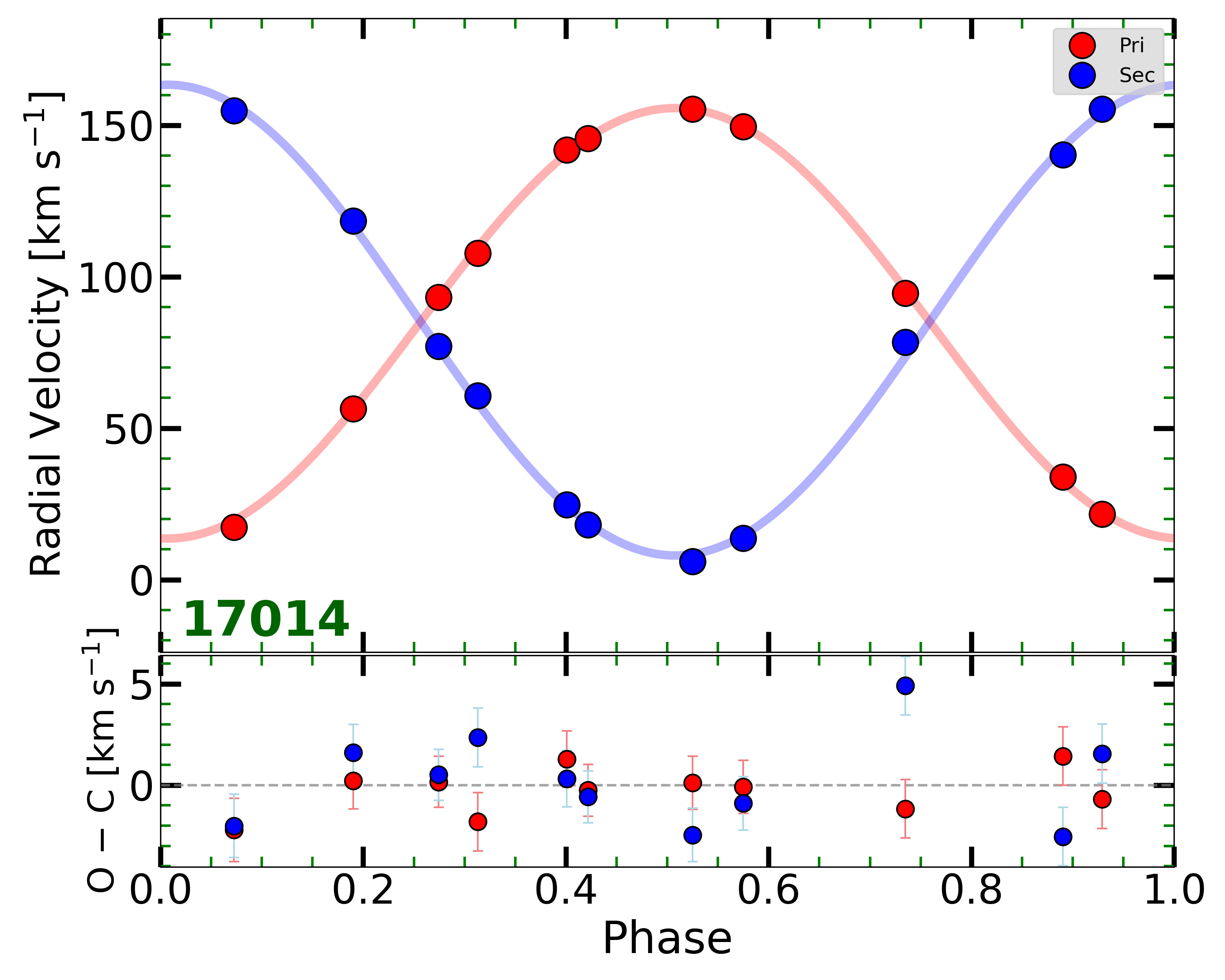}\\
\caption{Observed and modelled radial velocity variations for five binary systems analyzed, along with the residuals. Red symbols represent the primary components, and blue symbols represent the secondary components of the systems. \textit{See text for further details.}}
\label{fig:RVs}
\end{figure*}

\begin{table*}
\centering
\caption{Orbital parameters of the five double-lined spectroscopic binary systems analyzed in this study. For each system, the following parameters are listed: $K_1$ and $K_2$ are the radial velocity semi-amplitudes of the primary and secondary components, $P$ is the orbital period, $e$ is the orbital eccentricity, $\omega$ is the argument of periastron, $\tau$ is the time of periastron passage, and $\gamma$ is the systemic radial velocity}
\label{tab:binaries}
\begin{tabular}{lccccccc}
\hline
ID      & $T_0$           & $P_{\text{orb}}$ & $e$ & $\omega$ & $V_{\gamma}$ & $K_1$        & $K_2$ \\
 PKM    & (HJD - 2400000) & (days)           &     & (deg)    & (km s$^{-1}$)& (km s$^{-1}$)& (km s$^{-1}$) \\
\hline
5002 & $55915.2652 \pm 0.0115$ & $27.8665  \pm 0.0006$ & $0.589 \pm 0.004$ & $320 \pm 6$& $83.16 \pm 0.09$ & $62.02 \pm 0.21$ & $62.94 \pm 0.21$\\
17003 & $55792.2536 \pm 0.0028$ & $2.86764 \pm 0.00001$ & $0.188 \pm 0.001$ & $276$ & $84.79 \pm 0.09$ & $96.91 \pm 0.16$ & $113.45 \pm 0.16$ \\
24005 & $57436.4102 \pm 0.2083$ & $3.86981 \pm 0.00001$ & $0.004 \pm 0.001$ & $282 \pm 19$ & $83.74 \pm 0.07$ & $86.98 \pm 0.15$ & $88.27 \pm 0.15$ \\
17014 & $56179.9415 \pm 0.9155$ & $2.37263 \pm 0.00002$ & $0.007 \pm 0.005$ & $161 \pm 139$ & $85.13 \pm 0.30$ & $71.05 \pm 0.62$ & $77.70 \pm 0.61$ \\
9004 & $57479.1345 \pm 0.2039$ & $73.1326 \pm 0.0082$ & $0.534 \pm 0.009$ & $355 \pm 1$ & $84.74 \pm 0.22$ & $30.54 \pm 0.42$ & $30.44 \pm 0.58$ \\
\hline
\end{tabular}
\newline
\end{table*}

Among the five double-lined spectroscopic binary systems targeted in this study, two systems—WOCS~5002 and WOCS~17003—exhibit well-defined primary and secondary eclipses in their light curves. 
These systems were observed by TESS in multiple sectors (Sectors 7, 34, 61, and 88), providing high-cadence photometric data ideal for detailed eclipse modelling.
We extracted 2-minute cadence light curves from the TESS target pixel files using the \texttt{Lightkurve} package, applying custom detrending procedures to mitigate instrumental systematics and background contamination. Segments of the light curves containing complete eclipse events were carefully selected to ensure adequate phase coverage. In parallel, light curves from TGLC archive were used to validate our reductions and to cross-check flux levels and eclipse depths.

The high temporal resolution and photometric precision of the TESS data enabled us to robustly constrain key geometric and radiative parameters of the two binary systems, including orbital inclination, fractional radii, eccentricity, argument of periastron, and surface brightness ratio. These quantities, obtained from the TESS light curve modelling, serve as essential inputs for the subsequent SED and RV joint analyses. Notably, the strong eccentricity in WOCS~5002 ($e \sim 0.6$) introduces measurable asymmetries in eclipse timing and duration, providing additional constraints on orbital orientation.

\begin{table}
\scriptsize
	\begin{center}
		\caption{Obtained fundamental parameters for two eclipsing systems (WOCS 5002 and WOCS 17003) in NGC 2506 based on simultaneous LC and RV solutions. The standard errors 1$\sigma$ in the last digit are given in parentheses.} \label{tab:LCRV_results}
		\begin{tabular}{lll}
			\hline
			Parameter                                           &  WOCS 5002          & WOCS 17003          \\
			\hline 
			Initial epoch, T$_{\rm 0}$   (day)                  & 58506.199(76)  & 55792.223(11)      \\
			Period, P  (day)                                    & 27.86865(36)   & 2.8676185(95)        \\
            Geometric parameters:                               &                &                         \\
			Inclination, i ${({^\circ})}$                       & 85.5(1)        & 79.7(2)               \\
			Eccentricity, e                                     & 0.617(3)       & 0.174(6)              \\
			Argument of periapse, $\omega$ ${({^\circ})}$       & 315(6)         & 104(3)                \\
			  $\Omega _{1}$                                       & 21.42(21)      & 6.53(4)            \\
            $\Omega _{2}$                                       & 26.44(29)      & 8.98(5)               \\
            Fractional radii of pri.                            &                &                         \\ 
            $R_1/a$                                             & 0.0527(4)      & 0.1845(10)              \\
            Fractional radii of sec.                            &                &                          \\ 
            $R_2/a$                                             & 0.0427(6)      & 0.1113(7)              \\
            Radiative parameters:                               &                &                         \\ 
            T$_{\rm eff, \,1}$  (K)                             & 5523(20)       & 6894                   \\
            T$_{\rm eff, \,2}$  (K)                             & 6860           & 56820(35)               \\
            Albedo$^*$ ($A_1,A_2$)                              & 0.6, 0.6       & 0.6, 0.6                \\
            Gravity brightening$^*$ ($g_1, g_2$)                & 0.32, 0.32     & 0.32, 0.32              \\
            Light ratios                                        &                &                         \\
            $(\frac{l_1}{l_{\rm total}})$ (\%)                  & 20             & 74                    \\
            $(\frac{l_2}{l_{\rm total}})$ (\%)                  & 80             & 26                    \\
            \hline 
			Mass, $M_{1}$ ($\rm{M_{\odot}}$)                    & 1.512(18)      & 1.491(33)                        \\
			Radius, $R_{1}$  ($\rm{R{\odot}}$)                  & 2.938(115)     & 2.198(48)                      \\
			Luminosity, $L_1$ ($\rm{L_{\odot}})$                & 7.2(9)         & 9.80(97)                        \\
			Surface gravity, $\log g_1$ (cgs)                   & 3.682          & 3.928                              \\
			Mass, $M_{2}$ ($\rm{M_{\odot}}$)                    & 1.490(18)      & 1.273(24)                         \\
			Radius, $R_{2}$  ($\rm{R_{\odot}}$)                 & 2.380(93)      & 1.324(29)                          \\
			Luminosity, $L_{2}$ ($\rm{L_{\odot}})$              & 11.3(1.5)      & 3.40(34)                           \\
			Surface gravity, $\log g_{2}$ (cgs)                 & 3.858          & 4.300                                \\
            \hline 
		\end{tabular}
	\end{center}

\end{table}

\begin{figure*}
\centering
\includegraphics[width=0.4\linewidth]{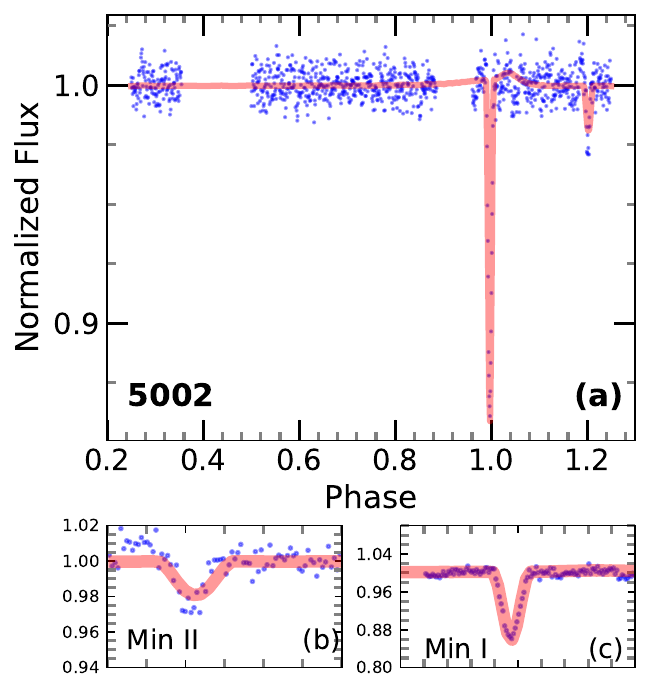}
\includegraphics[width=0.4\linewidth]{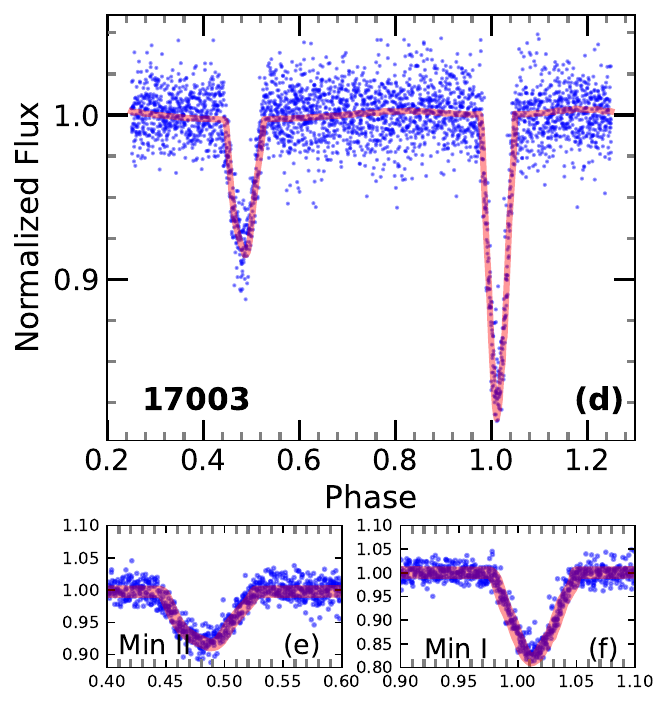}
\caption{Phase-folded TESS light curves (points) and best-fitting binary star models (solid lines) for WOCS~5002 and WOCS~17003. Panels (a) and (d) show the full light curves of the two systems folded with their orbital periods. Panels (b) and (c) show zoom-ins on the secondary and primary eclipses of WOCS~5002, respectively, while panels (e) and (f) show the same for WOCS~17003. The zoomed panels highlight the excellent agreement between the observed TESS photometry and the model fits near the eclipse minima.} \label{fig:LCs}
\end{figure*}

The light curve modelling was carried out using the \textsc{Phoebe} code \citep{Prsa2005}, which solves the binary geometry and surface flux distributions in a physically self-consistent manner. Model parameters such as orbital period, eccentricity, argument of periastron, inclination angle, fractional radii, surface brightness ratio, and third light contribution were simultaneously fitted. Initial parameter constraints were adopted from prior radial velocity solutions and Gaia DR3 astrometric data. 
Among the analysed systems, WOCS~5002 stands out due to its high orbital eccentricity ($e \sim 0.6$), which introduces additional complexity in interpreting eclipse depths and timings. In contrast, WOCS~17003 exhibits a much shorter period and a relatively low eccentricity, consistent with that of a typical close detached binary. The modelling yielded precise values for key parameters such as relative radii, temperature ratios, and orbital inclinations. These parameters serve as critical inputs --- especially the radial velocity semi-amplitudes (K$_1$ and K$_2$, commonly referred to as K velocities), eccentricities, and, in two cases, the orbital inclination angle --- for subsequent SED fitting and isochrone-based age analyses.
The best-fitting models for each system are summarised in Table~\ref{tab:binaries} and Table~\ref{tab:LCRV_results}, and the corresponding light curve fits are shown in Figure~\ref{fig:LCs}. To clearly show the agreement between the observations and models, both primary and secondary minima are plotted separately in the figure.

For all five SB2 systems selected in this study, we performed detailed radial velocity modelling using a custom-built MCMC-based pipeline developed specifically for this project. This approach enabled robust determination of orbital parameters including the systemic velocity, semi-amplitudes, eccentricity, and argument of periastron, along with realistic and uniform uncertainty estimates. 
The consistency of the solutions obtained from radial velocity observations for five systems is shown in Fig.~\ref{fig:RVs}. The derived K velocities and eccentricities served as strong priors in the subsequent light curve and SED analyses. In particular, for the two systems that exhibit eclipses (WOCS~5002 and WOCS~17003), the combination of high-precision radial velocity measurements and TESS photometry allowed for the determination of the stellar masses, radii, inclination angles, and temperature ratios with high accuracy. The consistency between the independently derived RV+LC solutions and subsequent SED fits underscores the reliability of the inferred parameters, and provides a solid foundation for constraining the evolutionary status of the component stars.

A detailed comparison with the orbital parameters derived by \citet{Knudstrup2020} for the eclipsing systems WOCS~5002 and WOCS~17003 reveals a high degree of consistency in the reported eccentricities and arguments of periastron, with differences remaining well within the stated uncertainties. Notably, while \citet{Knudstrup2020} employed the \textsc{JKTEBOP} and \textsc{ellc} light curve modeling codes, we used the \textsc{PHOEBE} software, which is explicitly optimized for eclipsing binary modeling based on the Wilson--Devinney framework. The use of different modeling approaches for similar systems provides a valuable cross-check. In addition, our analysis incorporates a more extensive set of \textit{TESS} observations and a unified RV+LC+SED modeling strategy. Following the full modeling described in Section~4, we find that the component masses and radii differ from those reported by \citet{Knudstrup2020} by approximately 0.5\% in mass and 2.5\% to 9.5\% in radius, respectively. While these fractional differences are modest in absolute terms, they exceed the typical uncertainties (1–3\%) quoted in both studies for some components, particularly in radius. This highlights how small deviations in light curve modeling—especially in the treatment of eclipse shapes and inclination angle—can propagate into meaningful differences in inferred stellar radii and, by extension, age estimates.

\section{Joint Sed Fitting for the Five Binary SEDs}
\label{sec:SEDs}

To find the system age of, and the distance to, NGC 2506 we jointly fit the SED data for the five binaries.  There are typically 24 SED points at different wavelengths, spanning 0.35 $\mu$m to 4.6 $\mu$m, available on VizieR SED \citep{ochsenbein00} for each of the five binaries.  Specifically, there are 118 total SED points.  The input data to the fitting code consist of these 118 SED points, 10 K-velocity values (found from fitting the RV data -- see Fig.~\ref{fig:RVs}), five eccentricity values (also found from fits to the RV data), and constraints on the orbital inclination angle for the two eclipsing binaries (derived from the eclipsing lightcurves)  There are 18 fitted parameters: 10 stellar masses, five orbital inclination angles, and a common system age, distance and interstellar extinction. 

The fit was carried out via a basic MCMC code (see, e.g., \citet{Yakut2025}).  At each link in the MCMC chain we have the stellar masses, the cluster age, and the cluster distance.  We use MIST evolution tracks \citep{Choi2016,dotter16,paxton11,paxton15,paxton19}  to infer the stellar radii and effective temperatures.  In turn, from these we can compute the model SEDs, with the use of the \cite{castelli03} model atmospheres.  The masses, orbital periods, eccentricities, and inclination angles are used to predict the K velocities.  And, these are compared to the measured K velocities to find their contribution to $\chi^2$.  The main runs were made for a nominal cluster metallicity of [Fe/H] = -0.30 (see Table~\ref{tab:ngc2506_sedresults}).  We ran numerous chains of 300 million links each. The reported parameter values and the corresponding uncertainties are taken to be the median values of the posteriors and their rms values, respectively.

\begin{figure*}
\centering
\begin{minipage}[b]{0.45\textwidth}
\centering
\includegraphics[width=\textwidth]{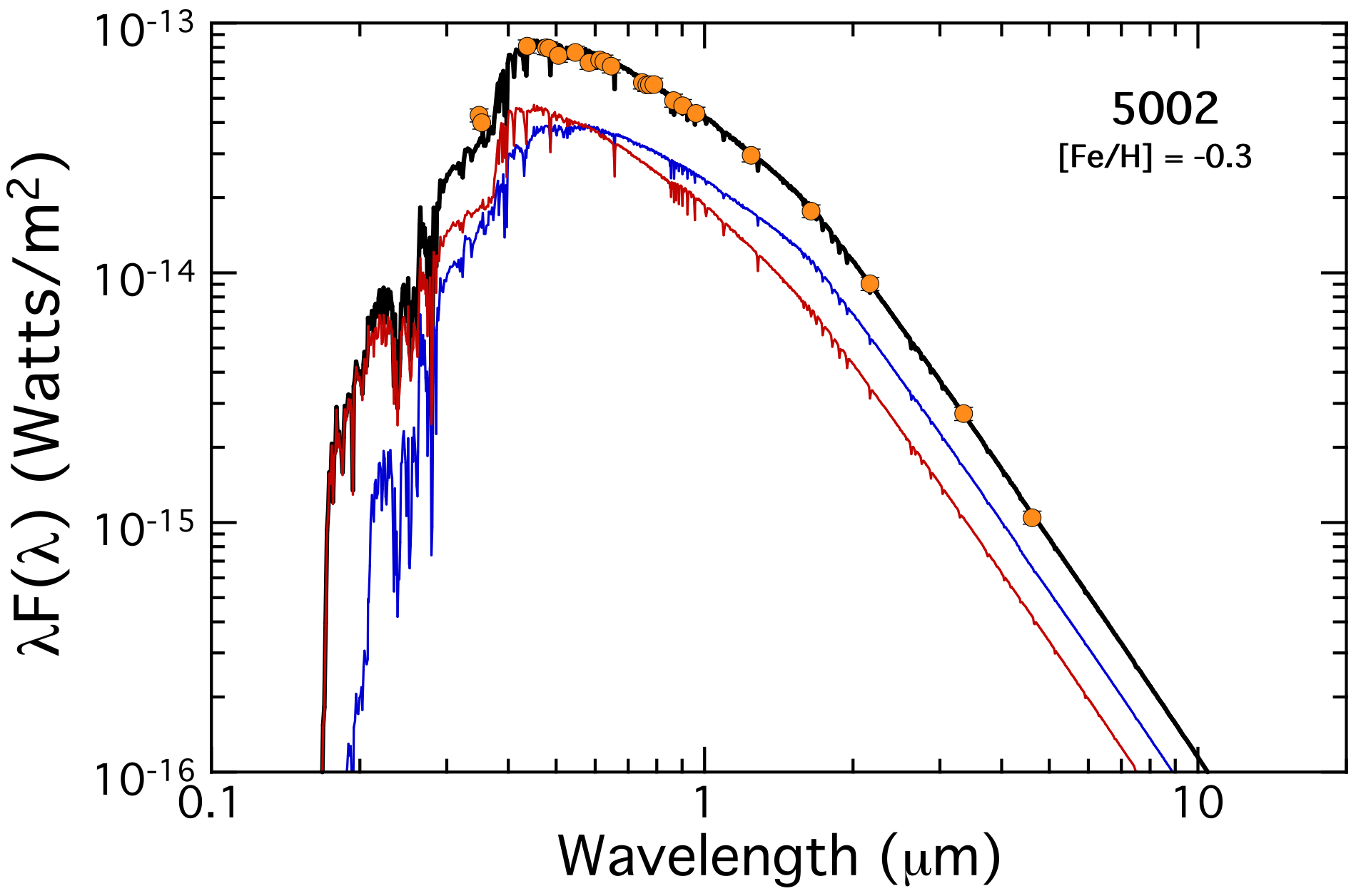}
\end{minipage}
\hspace{0.05\textwidth}
\begin{minipage}[b]{0.45\textwidth}
\centering
\includegraphics[width=\textwidth]{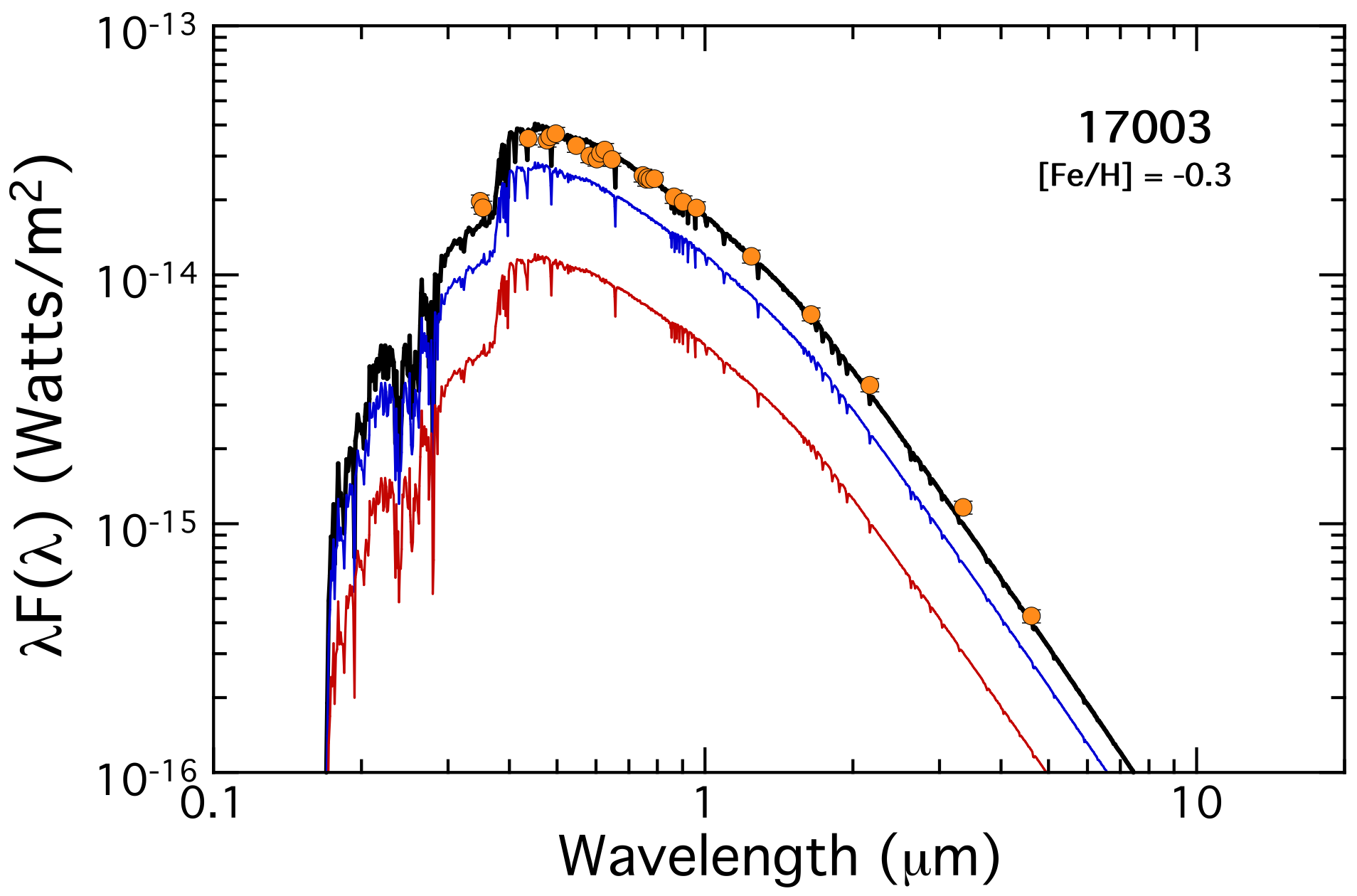}
\end{minipage}

\vspace{0.5em}  

\begin{minipage}[b]{0.45\textwidth}
\centering
\includegraphics[width=\textwidth]{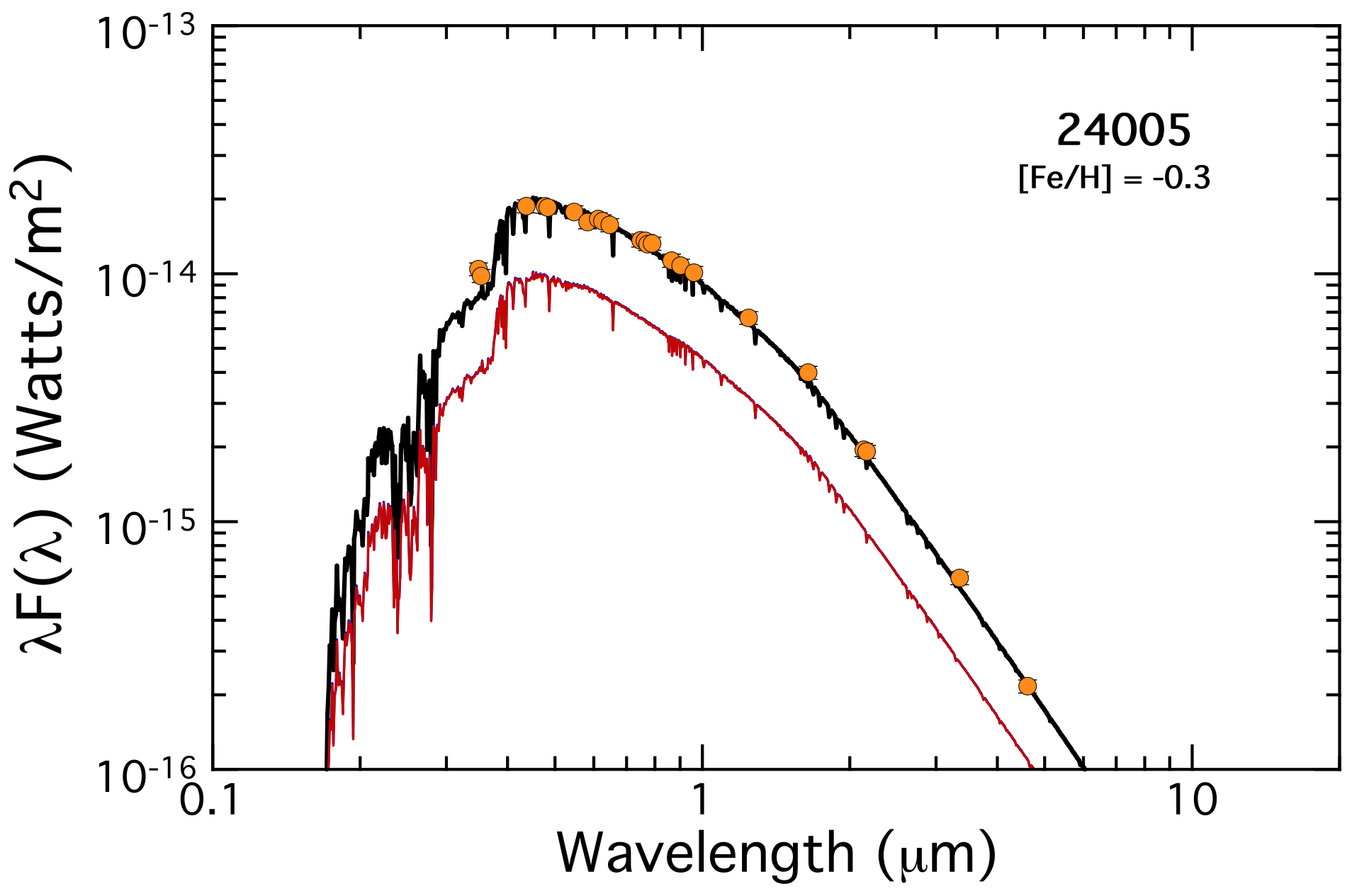}
\end{minipage}
\hspace{0.05\textwidth}
\begin{minipage}[b]{0.45\textwidth}
\centering
\includegraphics[width=\textwidth]{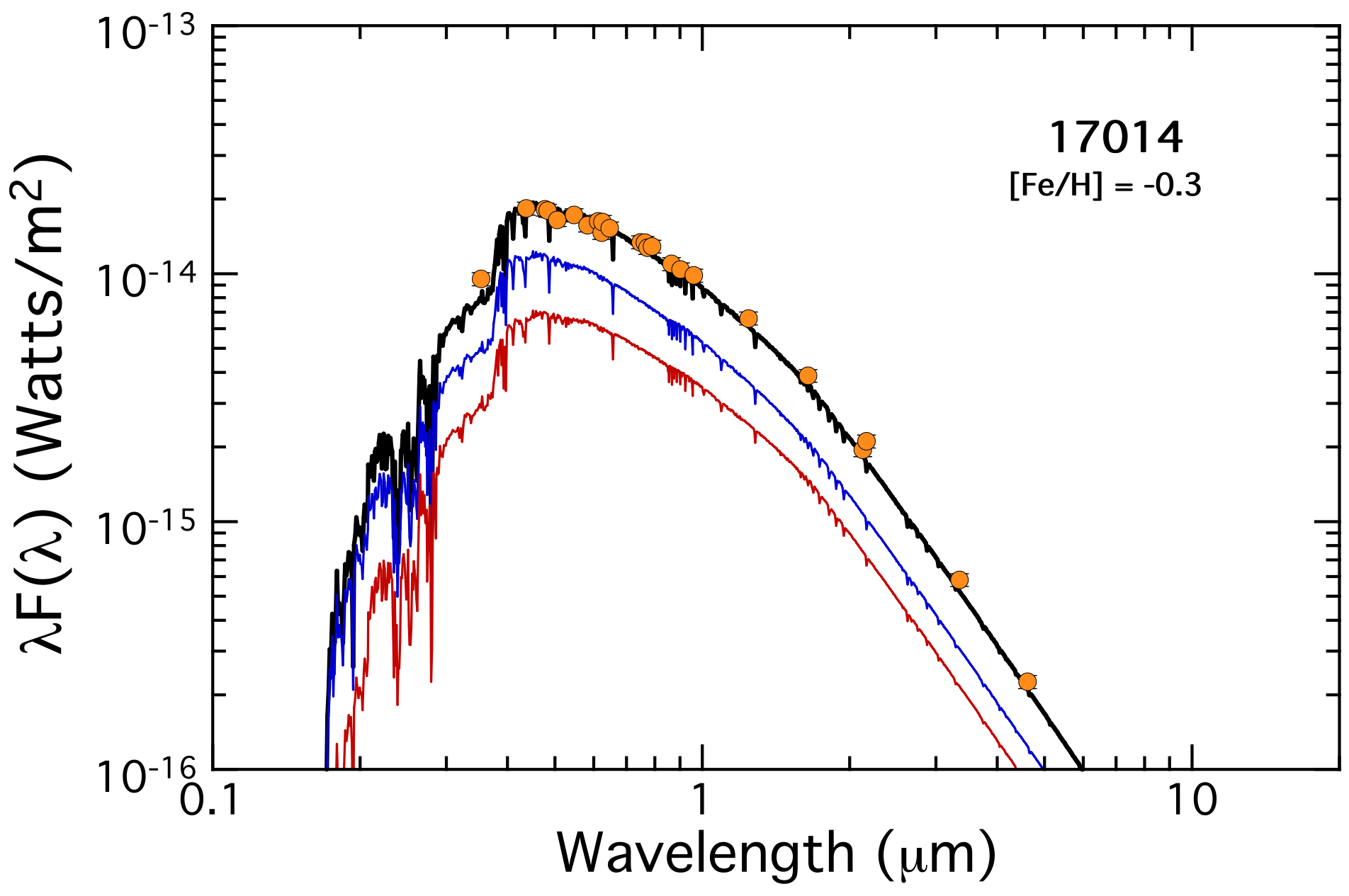}
\end{minipage}

\vspace{0.5em}  

\begin{minipage}[b]{0.45\textwidth}
\centering
\includegraphics[width=\textwidth]{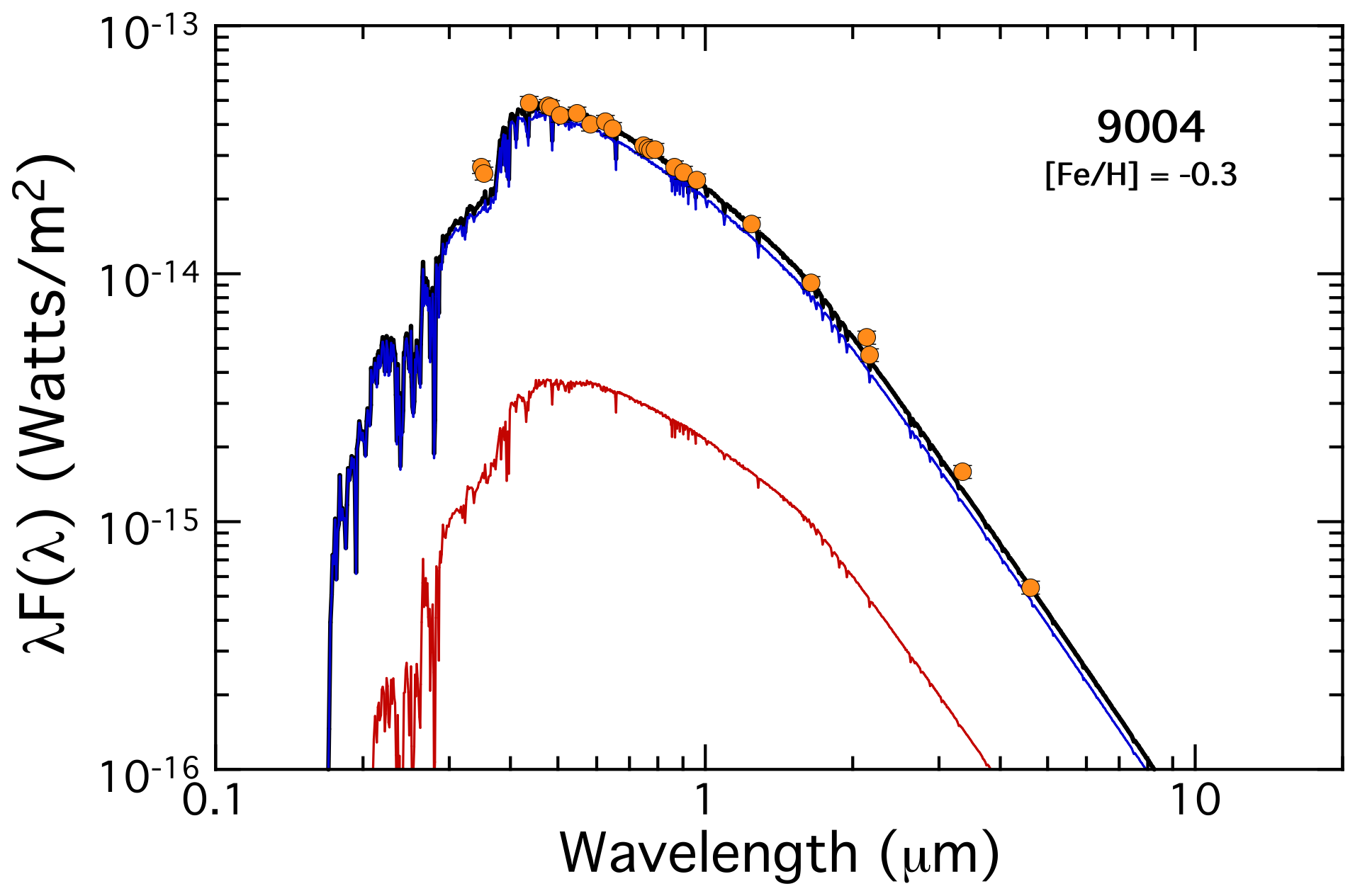}
\end{minipage}

\caption{We performed a joint SED fit for five binary systems, assuming a common age, distance, and extinction ($A_V$). The models were computed for a metallicity of [Fe/H] = –0.30. Figure~\ref{fig:age_Fe} illustrates how the derived age depends on metallicity. In the figure, blue and red lines represent the best-fit models for the individual components of each binary, while the black line corresponds to the combined (composite) SED. Observational data are shown as orange circles with associated error bars.}
\label{fig:seds}
\end{figure*}

\begin{figure}
\centering
\includegraphics[width=0.99\columnwidth]{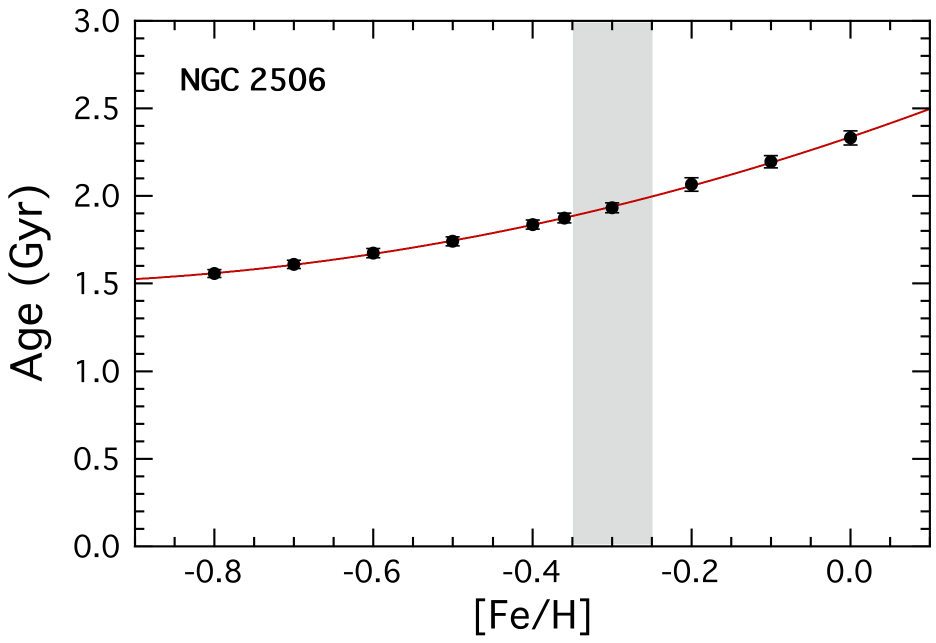}
\caption{Age of NGC 2506 vs.~the assumed metallicity [Fe/H]. The same type of SED fits used to produce Table \ref{tab:ngc2506_sedresults} and Fig.~\ref{fig:seds} was carried out for  additional metallicites. We find a parabolic relation between the age of NGC 2506 and metallicity given in Sec.~\ref{sec:SEDs}.}
\label{fig:age_Fe}
\end{figure}

The results of the SED fitting are summarized in Table~\ref{tab:ngc2506_sedresults} and displayed in Fig.~\ref{fig:seds}.  The age of NGC 2506 comes out to be $1.94 \pm 0.03$ Gyr, the photometric distance is $3189 \pm 53$ pc, and the extinction to the cluster is $A_V = 0.21 \pm 0.04$.  This age is a systematic function of the adopted metallicity.  Therefore, we ran ten SED fits, each for a different fixed metallicity in the range of $-0.8 < [{\rm Fe/H}] < 0.0$.  The results are plotted in Fig.~\ref{fig:age_Fe},  The fitted parabolic curve to the age-metallicity points can be expressed as
$${\rm age} = 1.94 + 1.1([{\rm Fe/H}]+0.3)+ 0.7([{\rm Fe/H}]+0.3)^2~{\rm Gyr}$$
where the pivot point is taken to be [Fe/H] = -0.30, i.e., the nominal metallicity.  In Sect.~\ref{sec:intro} we estimated the uncertainty in [Fe/H] to be 0.05.  If we use the above expression for the age to evaluate the uncertainties in the age, we find
$${\rm age} = 1.94 \pm 0.03 \pm 0.05~{\rm Gyr}$$
where the first error bar is the statistical uncertainty and the second is due to the uncertainty in the metallicity.  We do not attempt to fit for the metallicity as another free parameter.  The reason is that, while the age does systematically depend on metallicity, $\chi^2$ changes only very slowly with this parameter, and the metallicity is better determined via high-resolution spectroscopy.  We present the results as we do so that if, in the future, the metallicity determinations become even better, one can easily re-evaluate the age from this expression.

In summary, the age of the cluster is $1.94 \pm 0.03$ Gyr  (statistical error) or  $1.94 \pm 0.06$ Gyr,  where the latter error bar includes both the statistical and metallicity uncertainties.  The photometric distance is $3189 \pm 53$ pc which, as we shall see in the next section, agrees well with the distance inferred from Gaia.  Finally, the extinction to the cluster is estimated as $A_V = 0.21 \pm 0.04$. The mean of the measured values of reddening to NGC 2506 found in Table~\ref{tab:ngc2506_full} is $E(B-V) = 0.055 \pm 0.012$, where the given uncertainty is taken to be the rms scatter in the measurements.  If we use a nominal conversion between $E(B-V)$ and $A_V$ of 3.1 \citep{Fitzpatrick1999}, then this translates to $A_V = 0.17 \pm 0.04$, which is in good agreement with the extinction we find from the SED fitting.

Finally, we mention an issue that the SED fitting method we use here has to confront, namely being alert to possible  not fully resolved third stars, bound or unbound.  We have several ways of combating this. First, the stellar origin of an SED point is distinguishable in the visible between stars just a couple of arc sec apart.  They are still readily separable into the 2MASS NIR region (PSF = $2.1''$). However, as we progress through the WISE bands, the angular PSFs are $6.1''-6.3''$ for the W1 and W2 bands (none of our sources has flux measurements in either W3 or W4).  With regard to resolvable neighbor stars we have checked the Gaia archives for all stars within $20''$ of our target binary stars and having a $\Delta R_p < 4$.  In fact, two of the targets have no such contaminating stars, two have a single potentially contaminating star, and WOCS 9004 has four such neighbor stars. All the potentially contaminating stars have $\Delta R_p \gtrsim 3$, $B_p-R_p$ colors in the range of 0.77-0.93 (so not particularly red), and distances ranging from $6.3''$ to $8.6''$.  As suggested above, these should present no issue in the visible or 2MASS bands. In the WISE W1 and W2 bands, in principle, they could contribute up to 3\% of the flux in those bands (considering that the neighbors are all farther away than the W1 and W2 PSFs). Moreover, the Explanatory Supplement to the WISE All-Sky Data Release\footnote{https://wise2.ipac.caltech.edu/} suggests that ``the default fluxes are primarily based on PSF-fitting photometry, which includes modeled point spread functions and active/passive deblending algorithms.  While fixed-radius aperture fluxes (e.g., $8.25''$) are also available, they are not the default for most sources.’’  Therefore, assuming that the WISE fluxes we used were obtained with PSF-fitting, there should be no issue in this regard.

Second, we take the empirical point-to-point fluctuations in the VizieR SEDs at the magnitudes of the NGC 2506 stars to be $\sim$6\%.  Thus, statistically, with all 25 SED points taken together, we can determine overall fluxes to the percent level.  Therefore, if there is a third star in the system light, it is very likely to give us highly inconsistent results.  In fact, we had a number of binaries that we have analyzed, and discarded along the way, because we do not make any sense of the SED fits or find consistency with the input K velocities (e.g., WOCS 17013).  Third, in the event that three stars might mimic the parameters of a binary with incorrect masses, age, and/or distance, we rely on the fact that we have used five or more binaries in our studies of NGC 2506 and NGC 188  (10-12 stars), and that these other stars would strongly reveal the true distance and age, thereby exposing the binary contaminated by a third star.  In any case we try to be vigilant to this possibility.

\begin{table*}
\centering
\caption{Stellar parameters for 10 stars obtained from a joint SED analysis of five binary systems in NGC~2506. The IDs in uppercase denote the binary system labels, while lowercase letters (a, b) indicate the primary and secondary components, respectively. The cluster-level parameters derived jointly in the analysis are listed at the bottom of the table. All values were obtained assuming a metallicity of [Fe/H] = $-0.30$, which corresponds to $Z = 0.0150$, $X = 0.715$, and $Y = 0.270$ based on \citet{Choi2016} and \citet{Asplund2009}. The uncertainty in the derived distance is small because priors were set to match astrometric constraints (see Sect.~5). Additional analyses with varying [Fe/H] and distance priors were also explored.}
\label{tab:ngc2506_sedresults}
\begin{tabular}{lllllll}
\hline
PKM   & ID& Mass                & Radius           & Temperature   & Luminosity      & Inclination \\
      &   & (${M_{\odot}}$)     & (${R_{\odot}}$)  & ($K$)         & (${L_{\odot}}$) & (${{^\circ}}$) \\
\hline
5002  & Aa   & $1.518 \pm 0.009$ & $3.333 \pm 0.137$ & $6187 \pm 196$ & $14.7 \pm  0.78$ & $85.7 \pm  0.1$ \\
      & Ab   & $1.503 \pm 0.008$ & $2.518 \pm 0.160$ & $7125 \pm 184$ & $14.7 \pm  0.56$ & $85.7 \pm  0.1$ \\
17003 & Ba   & $1.475 \pm 0.007$ & $2.210 \pm 0.038$ & $6900 \pm  26$ & $10.0 \pm  0.27$ & $79.7 \pm  0.1$ \\
      & Bb   & $1.260 \pm 0.007$ & $1.421 \pm 0.014$ & $6985 \pm  18$ & $ 4.38 \pm  0.11$ & $79.7 \pm  0.1$ \\
24005 & Ca   & $1.215 \pm 0.011$ & $1.334 \pm 0.019$ & $6875 \pm  31$ & $ 3.61 \pm  0.18$ & $72.5 \pm  0.6$ \\
      & Cb   & $1.214 \pm 0.011$ & $1.332 \pm 0.019$ & $6871 \pm  31$ & $ 3.59 \pm  0.17$ & $72.5 \pm  0.6$ \\
17014 & Da   & $1.255 \pm 0.012$ & $1.412 \pm 0.024$ & $6975 \pm  29$ & $ 4.31 \pm  0.21$ & $44.1 \pm  0.4$ \\
      & Db   & $1.148 \pm 0.012$ & $1.209 \pm 0.021$ & $6670 \pm  39$ & $ 2.65 \pm  0.15$ & $44.1 \pm  0.4$ \\
9004  & Ea   & $1.510 \pm 0.009$ & $2.852 \pm 0.148$ & $6765 \pm 167$ & $15.3 \pm  0.38$ & $46.5 \pm  1.1$ \\
      & Eb   & $1.038 \pm 0.031$ & $1.027 \pm 0.048$ & $6306 \pm 103$ & $ 1.54 \pm  0.26$ & $46.5 \pm  1.1$ \\
\hline 
\multicolumn{3}{c}{{Age}} & \multicolumn{3}{c}{{Distance}} & {$A_V$} \\
\multicolumn{3}{c}{(Myr)} & \multicolumn{3}{c}{(pc)} & {(mag)} \\
\hline
\multicolumn{3}{c} {$1940 \pm 30$} & \multicolumn{3}{c}{$3189 \pm 53$} & {$0.21 \pm 0.04$} \\
\hline
\hline
\end{tabular}
\end{table*}

\section{Astrometric Analysis}
\label{sec:astrometric}

\begin{figure}
\centering
\includegraphics[width=0.90\linewidth]{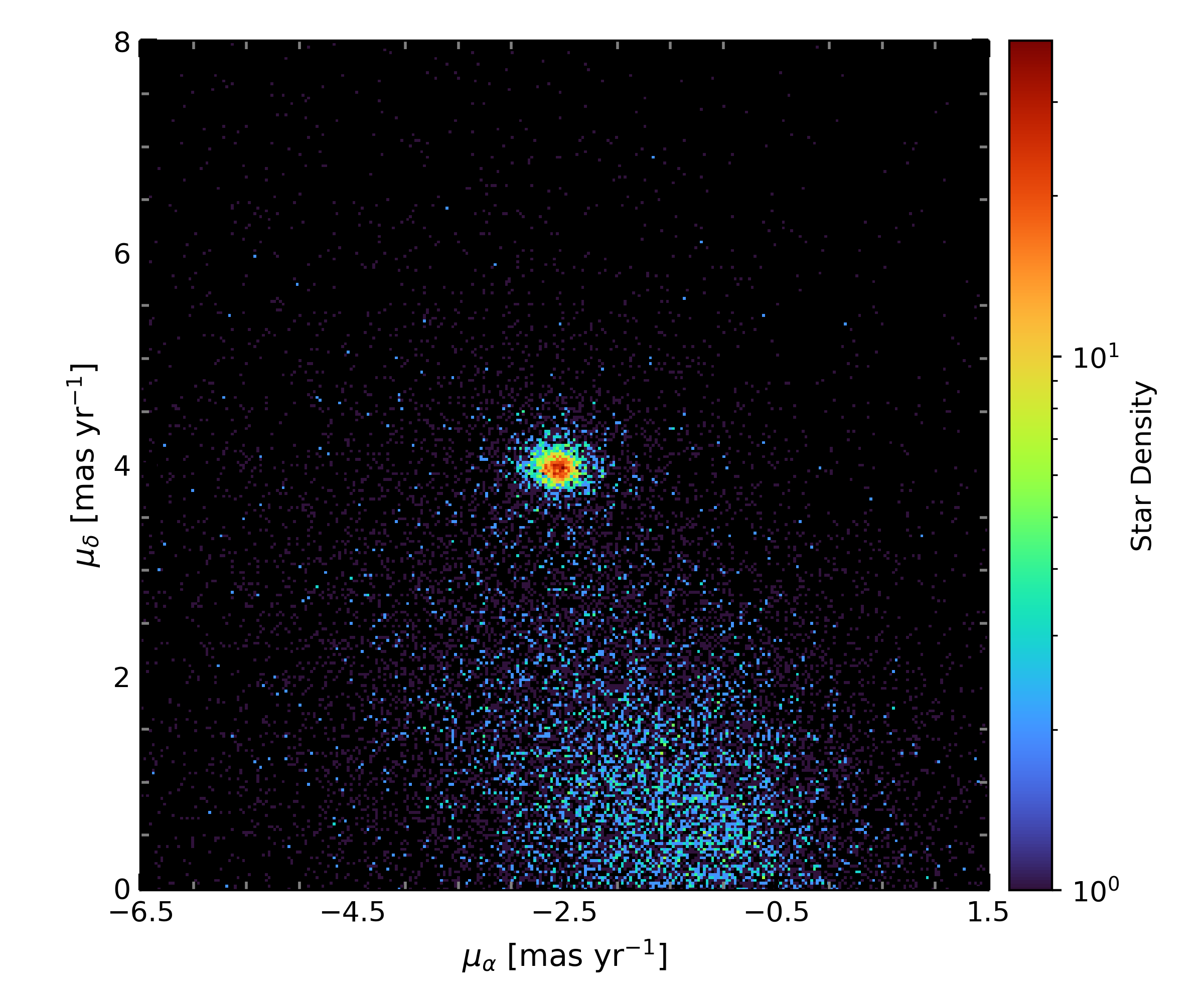}\\
\includegraphics[width=0.80\linewidth]{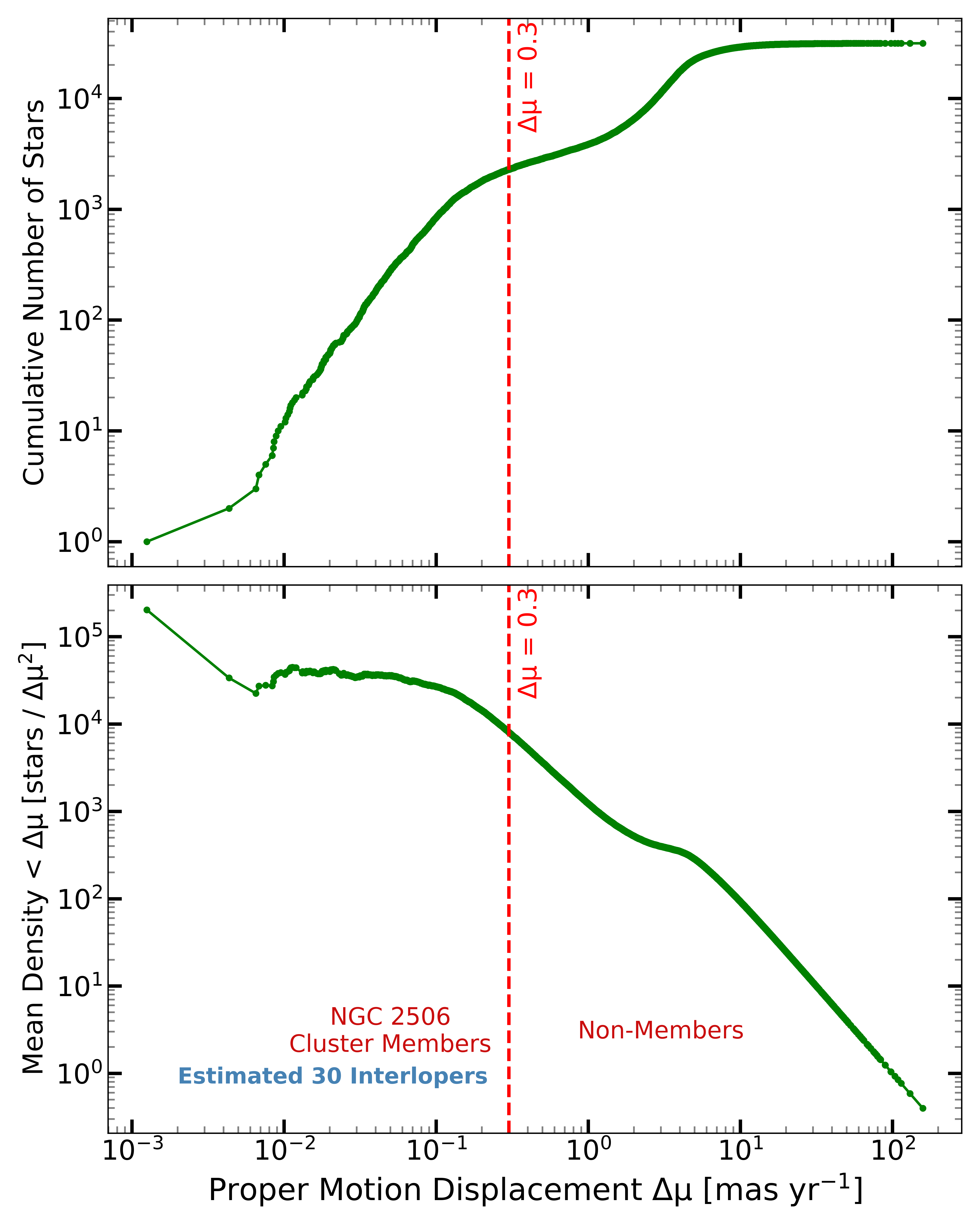}
\caption{\textbf{Top panel:} Proper motion density distribution of stars in the NGC 2506 field, based on Gaia DR3 data. 
\textbf{Middle panel:} Cumulative distribution of proper motion displacements $\Delta\mu$ from the cluster center  ($\mu_\alpha = -2.55 \pm 0.10$; $\mu_\delta = +3.97 \pm 0.09$). The red dashed vertical line, at the first major inflection point of the green curve marks the threshold value of $\Delta\mu = 0.3$ mas~yr$^{-1}$, used to distinguish cluster members from field stars.   \textbf{Bottom panel:} Mean cumulative density (number of stars per $\Delta \mu^2$) as a function of $\Delta\mu$. This plot helps visualize how steeply the density of stars drops right around the selected membership threshold, reinforcing the distinction between cluster members and non-members.}
\label{fig:pm_density_ngc2506}
\end{figure}

Reliable distance estimates are fundamental for deriving accurate stellar and cluster parameters. With the availability of high-precision Gaia astrometry and the probabilistic distance estimates published by \citet{BailerJones2021}, it is now possible to obtain robust distances for large stellar populations, including open clusters.
We identified likely members of NGC~2506 based on Gaia~DR3 proper motion and parallax information.
The selection was guided by the overdensity in the proper motion vector point diagram (see Figure~\ref{fig:pm_density_ngc2506}), focusing on stars within a narrow proper motion range centered on the cluster mean ($\mu_\alpha = -2.5$; $\mu_\delta = +4.0$), and applying a membership threshold of $\Delta\mu \leq 0.3$ mas~yr$^{-1}$ based on the overdensity in proper motion space.
To ensure data quality, we applied the following constraints: Gaia G-band magnitude $11 < G < 18$, RUWE~$\leq 1.4$, positive parallaxes ($\varpi > 0.05$\,mas), and relative parallax uncertainty $\sigma_\varpi / \varpi < 0.3$. This yielded a clean sample of 919 stars considered to be high-probability cluster members. A total of $\sim$30 of these stars are estimated to be field interlopers, based on their location in the proper motion and parallax space.

Figure~\ref{fig:pm_density_ngc2506} illustrates the selection process in detail. The top panel shows the two-dimensional density map of Gaia proper motions, where the concentrated overdensity marks the cluster's astrometric core. The middle panel presents the cumulative distribution of proper motion offsets $\Delta\mu$ relative to the cluster centroid, with a sharp inflection around $\Delta\mu = 0.3$ mas~yr$^{-1}$, which we adopt as the membership threshold. The bottom panel confirms this threshold by showing a steep drop in mean density beyond this value, indicating a relatively clean separation between cluster members and field contaminants. These diagnostics ensure that the selected sample used in our distance analysis is astrometrically coherent, with only minimal contamination from field stars.

We then extracted the distances for these stars directly from the Bailer-Jones catalogue. These distances are derived using a Bayesian inference scheme that accounts for the nonlinearity and noise properties of parallax measurements. The resulting distance distribution for the cluster members yields a median value of $3105 \pm 75$\,pc, where the uncertainty corresponds to the standard deviation across the sample. The standard error on the cluster distance is $\sim 2.4$\,pc, assuming uncorrelated individual uncertainties across the 919 members. Although the formal statistical error is small, we note that the physical depth of the cluster itself may contribute a comparable level of uncertainty. For instance, assuming an angular radius of $\sim$6$^{\prime}$  and a mean distance of $\sim$3100\,pc, the implied line-of-sight depth is about $\pm 5.5$\,pc. Therefore, the true uncertainty in the cluster’s distance centroid is likely dominated by intrinsic depth rather than statistical limitations. This value is in fairly good agreement with the independent estimate obtained from binary star SED analysis (see Table~\ref{tab:ngc2506_sedresults}), which yields $3189 \pm 53$\,pc, providing mutual consistency between astrometric and photometric approaches.

To assess the potential impact of field interlopers on the derived distance, we performed a Monte Carlo simulation following the same procedure as used for NGC~188 \citep{Yakut2025}. In the simulation, 30 field stars with a mean distance offset of 500\,pc and a larger dispersion ($\sim$360\,pc) were mixed with the cluster sample. The resulting median distance remained unchanged to within 3\,pc, indicating that contamination from this level of interlopers does not significantly bias the inferred distance for NGC~2506. In fact, our simulations show that there could even be up to 100 non-cluster interlopers in our sample before the onset of a significant distance shift.

\begin{figure}
\centering
\includegraphics[width=0.90\linewidth]{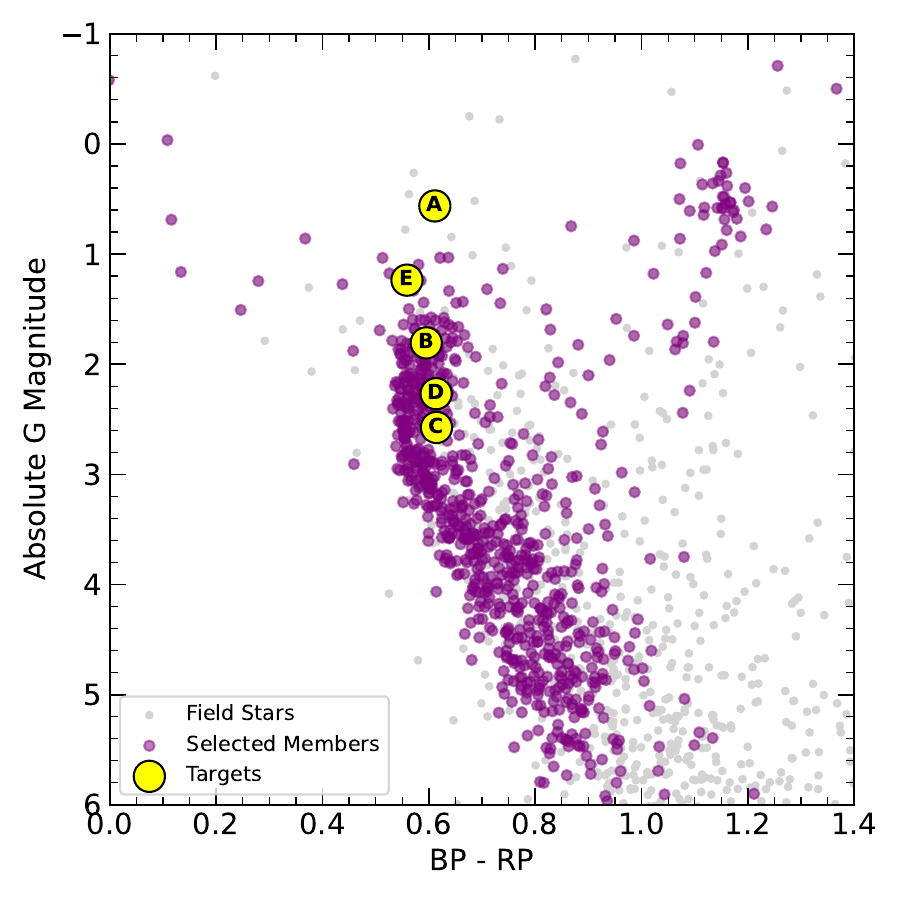} 
\caption{Hertzsprung–Russell diagram of NGC 2506 is plotted using Gaia photometry, where cluster members are shown in purple and the target binary systems are highlighted and labeled at their respective positions in yellow. The labels A–E correspond to the binary systems listed in Table~\ref{tab:ngc2506_sedresults}.}

\label{fig:gaiaDR3HR}
\end{figure}

We note that the RMS scatter in Bailer-Jones distances \citep{BailerJones2021} for NGC~2506 ($\sim$72\,pc) is significantly smaller than the $\sim$170\,pc found in our earlier analysis of NGC~188. This is likely due to the more compact structure of NGC~2506 and its better-defined membership in astrometric space, which reduces internal dispersion and minimizes projection effects. The selected distance estimates were used to compute absolute magnitudes and construct the Gaia-based Hertzsprung–Russell diagram (HRD), shown in Figure~\ref{fig:gaiaDR3HR}. The application of astrometric and photometric selection criteria substantially improved the clarity of the HRD, allowing a straightforward comparison with theoretical isochrones and providing an independent visualization on the SED-based age results. We adopted a metallicity of [Fe/H]~$= -0.30$ for the cluster in our subsequent analyses.

The resulting Gaia-based Hertzsprung–Russell diagram, shown in Figure~\ref{fig:gaiaDR3HR}, reflects the quality of the membership selection. The cluster sequence is well-defined and narrow, with minimal field contamination. Notably, the five binary systems studied in this work are clearly highlighted along the cluster main sequence and subgiant branches, demonstrating their consistency with the overall cluster population when accounting for their binary nature. The clean morphology of the HRD further supports the reliability of our astrometric filtering and adopted distance.

%%%%%%%%%%%%
%%%SECTION
%%%%%%%%%%%%
\section{Results and Conclusion}
\label{sec:results}

\begin{figure}
\centering
\includegraphics[width=0.465\textwidth]{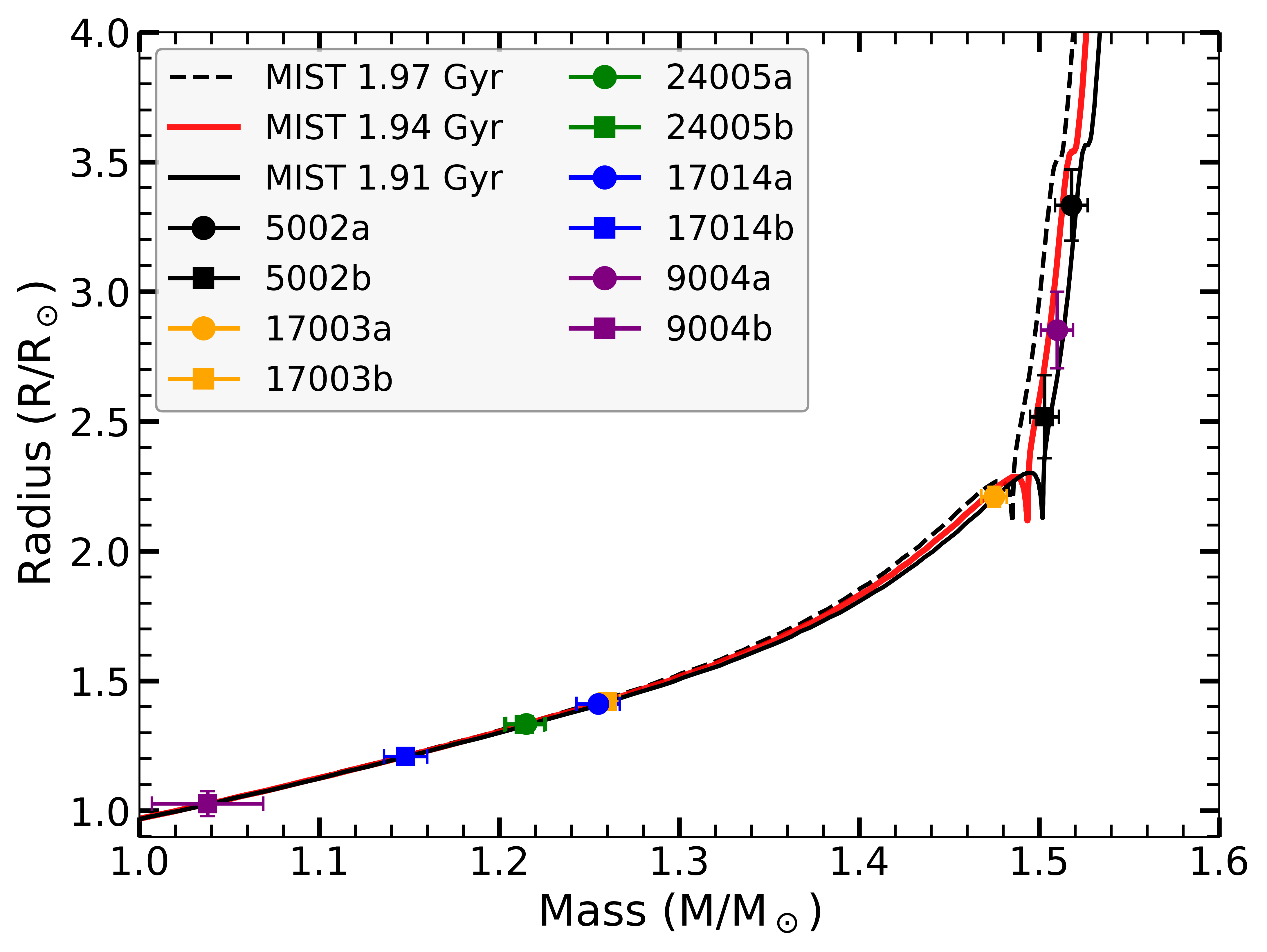}
\includegraphics[width=0.475\textwidth]{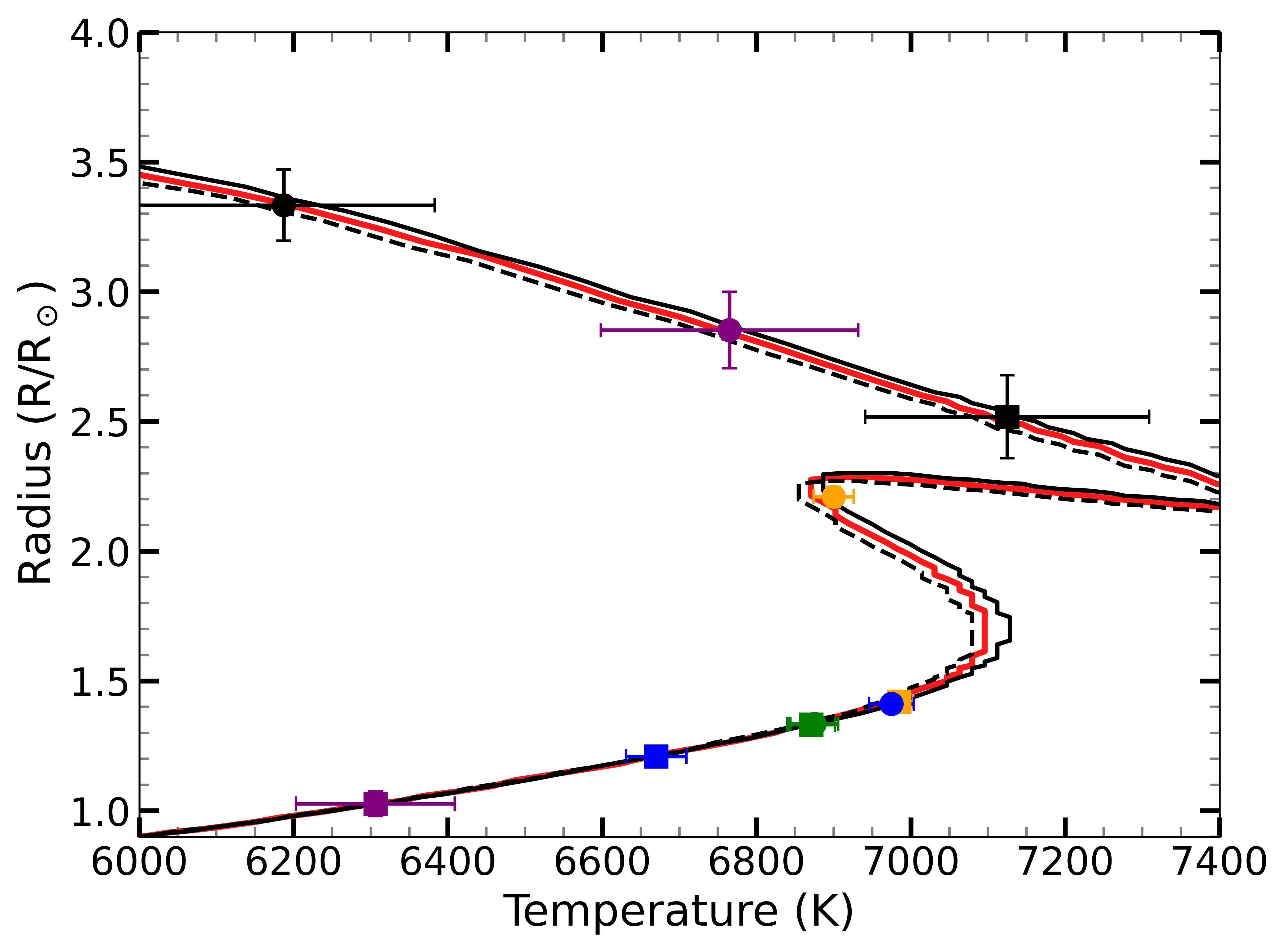}
\includegraphics[width=0.47\textwidth]{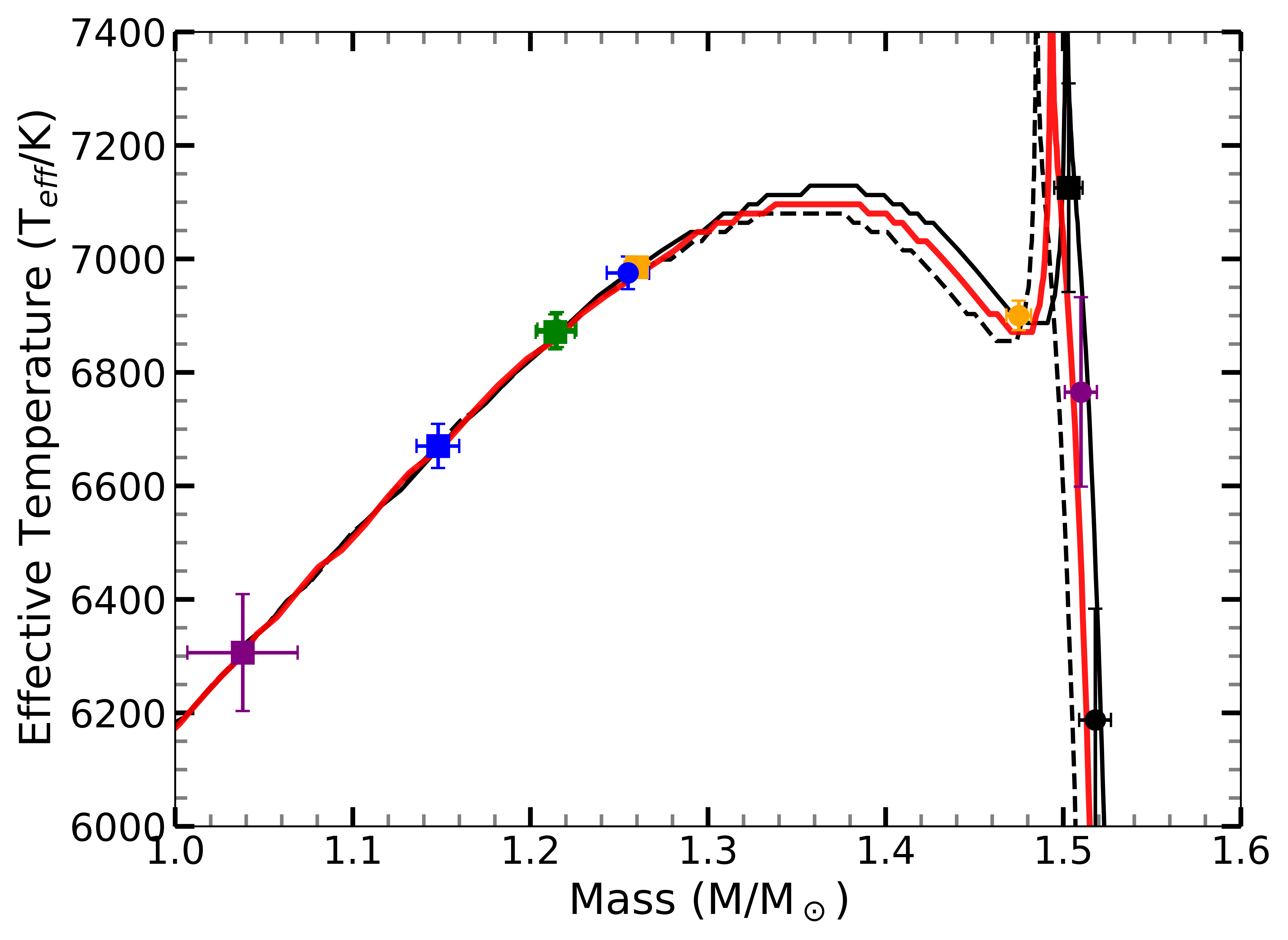} 
\caption{Mass–Radius, Radius–Temperature, and Temperature–Mass diagrams of five binary systems in the NGC~2506 cluster are presented, with the components of each system shown using circles (primaries) and squares (secondaries) in distinct colors. Stellar evolution isochrones from the MIST models (red) for an age of 1.94~Gyr are overplotted. To illustrate the uncertainty in the cluster age, additional isochrones for 1.91~Gyr (black solid line) and 1.97~Gyr (black dashed line) are also shown. The physical properties of the five target binaries are listed in Table~\ref{tab:ngc2506_sedresults}.} 
\label{fig:iso}
\end{figure}

Binary stars have long been recognized as fundamental astrophysical laboratories for deriving precise stellar parameters, particularly when they are double-lined and eclipsing. Traditionally, cluster studies have relied on one or two such systems per cluster to estimate key properties like distance and age \citep[e.g.,][]{Hensberge2000,Southworth2005,Yakut2009,Brogaard2012,Yakut2015}. While these earlier works laid important groundwork, they were often limited by sample size and observational constraints. More recently, however, the availability of high-precision photometric data (e.g., from TESS), high-resolution spectroscopy, and Gaia astrometry has opened a new era where multiple binaries within a cluster can be analyzed jointly. This enables the determination of not only individual stellar parameters but also global cluster properties with significantly improved precision and physical insight.

As summarized in Table~\ref{tab:ngc2506_full}, previous studies of NGC~2506 have reported a wide range of estimates for age (1.5--3.4~Gyr), metallicity ($\rm [Fe/H]$ from $-0.5$ to $-0.2$~dex), reddening ($E(B{-}V) = 0.04$--$0.10$), and distance. This scatter likely arises from differences in observational techniques, data quality, and the variety of methodologies adopted, including photometric isochrone fitting, spectroscopic analyses, and modeling of individual binaries.

In this work, we have conducted a detailed study of the intermediate-age open cluster NGC~2506 by modeling five well-characterized double-lined binary systems, two of which are eclipsing. These systems were selected based on their spectroscopic completeness, photometric quality, and astrometric membership probability. Each binary comprises components with masses ranging from approximately 1.0 to $1.5~M_{\odot}$, covering stages from near the ZAMS to the subgiant branch. For all ten stars, we performed a joint modeling of the radial velocity curves and SEDs, using synthetic spectra from \citet{castelli03} model atmospheres, and adopting MIST stellar evolution tracks \citep{Choi2016,dotter16} to derive the physical parameters of the components. This process was optimized using an MCMC framework, yielding robust estimates of temperatures, radii, and extinction values.

Complementing the binary modeling, we determined the astrometric distance using Gaia~DR3 parallaxes  and proper motions. We chose a narrow range of proper motions around the cluster center of $\Delta \mu < 0.3$ mas/year, as well as imposing a set of additional criteria as follows: photometric magnitude range $11 < G < 18$, RUWE~$\leq 1.4$, parallax~$>0.05$~mas, and fractional parallax error~$<30\%$. In all, we selected 919 high-confidence cluster members. 
For these stars, we adopted geometric distance estimates from the \citet{BailerJones2021} catalog, which accounts for parallax systematics using a Bayesian framework. The resulting median distance is $3105 \pm 75$~pc, where the uncertainty represents the standard deviation among the selected members. While the standard error of the mean is smaller ($\sim2.4$~pc), it likely underestimates the true distance uncertainty due to the cluster’s intrinsic depth. This result is in satisfactory agreement with the photometric distance of $3189 \pm 53$~pc derived from SED modeling (see Table~\ref{tab:ngc2506_sedresults}), demonstrating the consistency between astrometric and photometric approaches.

The age of the cluster was primarily determined through broadband SED fitting and radial velocities. Comparisons with MIST isochrones \citep{dotter16,Choi2016} were then performed purely as a sanity check, and as an alternative way to visualize the inferred parameters. Isochrones were computed for a range of ages and metallicities, with [Fe/H]~$= -0.30$ adopted as the baseline based on literature estimates and consistency with our SED analysis (see Figure~\ref{fig:iso}). While we did not treat [Fe/H] as a free parameter in the SED fitting, we propagated the observed rms scatter in the literature metallicity values (0.05~dex; see Table~\ref{tab:ngc2506_full}) into our age uncertainty estimates. The best-fit age from the SED-based analysis is $1.94 \pm 0.03$~Gyr (statistical), and only $\pm 0.06$ when adding in the uncertainty in metallicity. This value was then used to select a corresponding isochrone, which is compared against the observed stellar parameters in Figure~\ref{fig:iso}. The excellent match is, of course, mostly a sanity check since we were using MIST evolution tracks to do the fitting. To illustrate how the isochrones would change for a 0.1 Gyr offset, we  show two additional isochrones for 1.91 Gyr and 1.97 Gyr.

We note that, naturally, our results depend on the fidelity of the MIST evolution tracks. However, any theoretical model representing Nature is bound to have imperfections. To estimate uncertainties in our age determinations due to the choice of evolution tracks, we compared YaPSI/Y$^2$ \citep{Demarque2004,Spada2017} and PARSEC \citep{Bressan2012} tracks with those of MIST over a grid of 60 points (four masses from 1.2 to 1.5 M$_\odot$ and 15 radii from 2.0 to 3.5 R$_\odot$). For each grid point, we computed the fractional standard deviation of the three model ages. The average fractional standard deviation was found to be approximately 3.6\%, corresponding to an age uncertainty due to the models of about ±0.07 Gyr for NGC 2506’s age of 1.94 Gyr. This model-dependent uncertainty is comparable to other sources of error, such as statistical uncertainty in our SED fitting and uncertainty in chemical composition. We have also computed the mean fractional standard deviations over a wider model range than applies only to this problem, i.e., masses of 1.2-2.4 M$_\odot$ and radii from 2-8 R$_\odot$, and this comes out to be 6\%. While we do not provide a formal combined error budget for the age of NGC 2506, one should be aware of model limitations  for all age determinations using theoretical isochrones.

This study delivers the first homogeneous joint RV+SED analysis for five double-lined binary systems in NGC  2506, enabling accurate determinations of stellar masses, radii, and temperatures for ten well-characterized cluster members. Building upon the same methodology that we previously applied with success to NGC 188 \citep{Yakut2025}, we demonstrate that combining spectroscopic orbits with broadband SEDs and Gaia-based distance priors yields a physically grounded and internally consistent cluster age estimate. Unlike traditional isochrone-based approaches, which infer stellar masses indirectly from photometric loci, our method uses directly measured masses to constrain stellar evolution models. For NGC 2506, this leads to a tightly constrained cluster age of $1.94 \pm 0.03$,Gyr and a Gaia-independent distance  of $3189 \pm 53$,pc, significantly narrowing the wide range of literature values and helping reconcile previous discrepancies. The modeling framework also properly accounts for the binary nature of the sources, which, if ignored, could introduce systematic biases in inferred cluster parameters such as age and distance.

Beyond improving the properties of individual stellar systems, our results position NGC 2506 as a well-characterized intermediate-age open cluster suitable for testing stellar evolution models at subsolar metallicities. The mutual consistency across all five binaries, achieved without tuning isochrones or applying empirical offsets, demonstrates the self-consistency and reliability of the derived parameters. By adding NGC 2506 to the small but growing set of open clusters with precisely modeled SB2 systems, this study underscores the promise of joint RV+SED modeling as a complementary approach to traditional isochrone fitting, particularly for clusters with complex stellar populations or ambiguous color-magnitude morphologies. The methodology is scalable and readily applicable to other clusters with suitable spectroscopic binaries, offering a viable path toward precision stellar population studies in the Gaia era.

\section*{Acknowledgements}
We thank an anonymous referee for some very constructive suggestions for improving the manuscript. We thank Willie Torres (CfA) for prompting the discussion regarding potential third-light contamination in the SED analysis.
This work has made use of data from the European Space Agency (ESA) \textit{Gaia} mission (\url{https://www.cosmos.esa.int/gaia}), processed by the Gaia Data Processing and Analysis Consortium (DPAC; \url{https://www.cosmos.esa.int/web/gaia/dpac/consortium}). It also makes use of observations obtained with the Transiting Exoplanet Survey Satellite (\textit{TESS}) mission, funded by the NASA Explorer Program and publicly available through the Mikulski Archive for Space Telescopes (MAST). This research has made use of the SIMBAD and VizieR catalogues operated at the Centre de Données astronomiques de Strasbourg (CDS), and of NASA’s Astrophysics Data System (ADS) bibliographic services. This study was supported by the Scientific and Technological Research Council of T\"urkiye (TÜBİTAK, grant numbers 112T766, 117F188 and 2219). Numerical computations were performed in part using the High Performance and Grid Computing Center (TRUBA resources) provided by TÜBİTAK ULAKBİM. KY acknowledges the support of Churchill College, University of Cambridge, through a research fellowship.

%%%%%%%%%%%%%%%%%%%%%%%%%%%%%%%%%%%%%%%%%%%%%%%%%%
\section*{Data Availability}

The TESS observations used in the analysis of binary light curves are available online to the public through the Mikulski Space Telescope Archive (MAST). If desired, the data used in this study can be obtained from the MAST servers or requested from the authors.



\bibliographystyle{mnras}
\bibliography{ngc2506} % if your bibtex file is called example.bib

%\begin{thebibliography}{99}
%\end{thebibliography}

%%%%%%%%%%%%%%%%%%%%%%%%%%%%%%%%%%%%%%%%%%%%%%%%%%

% Don't change these lines
\bsp	% typesetting comment
\label{lastpage}
\end{document}